\documentclass[11pt]{article}

\usepackage{xr}
\makeatletter
\newcommand*{\addFileDependency}[1]{
  \typeout{(#1)}
  \@addtofilelist{#1}
  \IfFileExists{#1}{}{\typeout{No file #1.}}
}
\makeatother

\usepackage[english]{babel}
\usepackage[utf8x]{inputenc}
\usepackage[T1]{fontenc}
\usepackage{amsmath,amssymb}
\usepackage{algorithm}
\usepackage{mathtools}
\usepackage[noend]{algpseudocode}
\makeatletter
\def\BState{\State\hskip-\ALG@thistlm}
\makeatother

\usepackage[numbers,sort&compress]{natbib}
\usepackage[a4paper,top=3cm,bottom=2cm,left=3cm,right=3cm,marginparwidth=1.75cm]{geometry}
\usepackage{graphicx}
\usepackage{subcaption}
\usepackage{float}
\usepackage{accents}
\usepackage{fancybox}
\usepackage{geometry}
\allowdisplaybreaks
\title{Tuning acoustic impedance in load-bearing structures
}

\author{Sai Sharan Injeti$^\textrm{a}$, Paolo Celli$^\textrm{a,b}$, Kaushik Bhattacharya$^\textrm{a}$ and Chiara Daraio$^\textrm{a}$\footnote{Email: daraio@caltech.edu}}
\date{\scriptsize $^\textrm{a}$Department of Mechanical and Civil Engineering, California Institute of Technology, Pasadena, CA 91125, USA \\ $^\textrm{b}$Department of Civil Engineering, Stony Brook University, Stony Brook, NY 11794, USA}

\def\undertilde#1{\mathord{\vtop{\ialign{##\crcr
$\hfil\displaystyle{#1}\hfil$\crcr\noalign{\kern1.5pt\nointerlineskip}
$\hfil\tilde{}\hfil$\crcr\noalign{\kern1.5pt}}}}}
\usepackage{color}
\begin{document}
\maketitle
\begin{abstract}
\normalsize
\bigskip
Acoustic transparency is the capability of a medium to transmit mechanical waves to adjacent media, without scattering. This characteristic can be achieved by carefully engineering the acoustic impedance of the medium -- a combination of wave speed and density, to match that of the surroundings. Owing to the strong correlation between acoustic wave speed and static stiffness, it is challenging to design acoustically transparent materials in a fluid, while maintaining their high structural rigidity. In this work, we propose a method to design architected lattices with independent control of the elastic wave speed at a chosen frequency, the mass density, and the static stiffness, along a chosen loading direction. We provide a sensitivity analysis to optimize these properties with respect to design parameters of the structure, that include localized masses at specific positions. We demonstrate the method on five different periodic, three dimensional lattices, to calculate bounds on the longitudinal wave speed as a function of their density and stiffness. We then perform experiments on 3-D printed structures, to validate our numerical simulations. The tools developed in this work can be used to design lightweight and stiff materials with optimized acoustic impedance for a plethora of applications, including ultrasound imaging, wave filtering and waveguiding. 

%Acoustically transparent materials possess carefully engineered acoustic impedance -- a combination of sound speed in the material and its density. Several applications ranging from low frequency musical instruments to high frequency ultrasound imaging require matching acoustic impedance over media to attain maximum transmission of sound energy across interfaces. However, due to the strong correlation between acoustic wave speed in a material and its static stiffness, these materials often fail when structural rigidity is necessary together with acoustic transparency. In this work, we propose a method to design architected lattices with independent control over the elastic wave speed of propagation at a chosen frequency, together with its mass density, and static stiffness along a chosen loading direction. We provide a sensitivity analysis to optimize these properties with respect to design parameters of the structure, that include localized masses at specific positions. We demonstrate the method on five different periodic, three dimensional lattices (modeled using frame elements), to calculate bounds on the longitudinal wave speed as a function of their density and stiffness. We then perform experiments on 3-D printed structures that show great agreement with our numerical simulations. The tools developed in this paper can be used to design lightweight and stiff structures with optimized acoustic impedance, frequency filters, delay lines, vibration absorbers and three dimensional waveguides.

\end{abstract}
\bigskip
\normalsize

\section{Introduction}
In acoustics, the transmission coefficient can be defined as the ratio of amplitude of the transmitted wave to that of the incident wave at an interface between two media~\cite{graff2012}. This ratio is unity when the acoustic impedances (product of density and wave speed) of the two media are identical~\cite{graff2012}, ensuring maximum transmission of elastic wave energy across the interface. This is of particular importance in the design of acoustically transparent materials, where the acoustic impedance of the material matches that of the surroundings, and produces negligible interference with the propagating acoustic waves. Traditionally, while engineering the acoustic impedance of a structure, there is not much freedom in designing its static stiffness. This is because the static stiffness (along the direction of wave propagation) varies with the square of the elastic wave speed, making it difficult to design structures with high stiffness but low or moderate wave speeds. There are several applications where the mechanical rigidity of these acoustically transparent media is critical, including underwater sonar windows, acoustic imaging using transducer arrays, medical ultrasonography, hydrophone casings and other support structures for polymer transducers~\cite{thompson1982, thompson1990, anderson2008, chen2009, salzer2012}. In this work, we present a design approach that decouples the static stiffness of a structure from its elastic wave speed. We apply this method to architected lattices featuring point masses at specific locations, which act as localized scatterers or resonating elements in the dynamic response of these structures.

Architected solids are a class of materials whose macroscopic properties stem from a carefully-engineered \emph{mesostructure}, whose characteristic lengthscale is in between the atomistic one (that dictates the behavior of the constituent material) and the overall size of the system~\cite{Fleck2010}. Commonly, these mesostructures are obtained by spatially repeating a unit cell or repetitive volume element (RVE) that comprises peculiar spatial arrangements of material phases and voids. An engineering-relevant subclass of architected solids are lattice structures, i.e., networks of simple structural elements like trusses~\cite{Deshpande2001_2}, shells~\cite{Vidyasagar2018} or plates~\cite{Berger2017}, that show unusual combinations of mechanical properties like high strength and light weight \cite{Ashby2006}. Most early studies on lattice structures focused on their peculiar mechanics, e.g., on their ability to display mechanisms of inextensional deformation and states of self stress~\cite{Injeti2019,Hutchinson2006, bilal2017intrinsically}, and on their potential structural applications, e.g., as cores of sandwich panels~\cite{Evans2001, Deshpande2001}. Lattice structures with increasingly complex architectures can now be additively fabricated in many different materials, from polymers and metals~\cite{Injeti2019, Pham2019}, to ceramics~\cite{Meza2014}, composites~\cite{Raney2018} and cementitious materials~\cite{Moini2018}, and with characteristic lengthscales reaching down to nanometers~\cite{Bauer2017}. 

Lattice structures are particularly appealing for their response to dynamic loads, since their complex architectures give way to peculiar dispersion properties~\cite{Phani2017}. In the context of elastic wave propagation, periodic lattices can exhibit bandgaps, i.e., frequency ranges where waves are not allowed to propagate~\cite{Phani2006, Gonella2008, Liebold2014, Kroedel2014, Wang2015}. Beyond this well-known attribute, lattices also exhibit spatial wave manipulation capabilities, e.g., wave anisotropy or directionality~\cite{Casadei2013, Celli2014, Bayat2017}, negative refraction~\cite{Tallarico2017} and topologically-protected backscattering-free waveguiding~\cite{Vila2017}, stemming from intrinsic or carefully-chosen unit cell symmetries that yield mode-rich dispersion relations. In acoustics, microlattice metamaterials immersed in a fluid behave as poroelastic media and interact with ultrasonic waves, leading to local resonance based bandgaps~\cite{Spadoni2014, Kroedel2016} and other wave manipulation effects such as wave focusing~\cite{Su2017}, with potential applications in the biomedical field~\cite{Jimenez2019}. When manufactured at the micro- or nano-scale, lattice architectures can also interact with electromagnetic waves and exhibit photonic gaps~\cite{Chernow2015}. These characteristics, coupled with their structural performance, make lattices appealing as multifunctional systems for mechanical, aerospace and biomedical applications.

A critical need to translate structured materials into engineering applications is the development of inverse design strategies, to obtain optimal architectures given desired specifications. %and to try and reach the limits in terms of properties that can be achieved. 
%This is a treatable problem for periodic structures, since it is sufficient to optimize the properties of a single cell by imposing adequate periodic boundary conditions. 
In dynamics, the dispersion properties of a periodic lattice are often derived from the analysis of a single cell, applying periodic boundary conditions of the Floquet-Bloch type, which impose a dependency on the wavenumber~\cite{Brillouin1953, Phani2006}. This approach simplifies the optimization process, and reduces it to the design of a single cell. Most optimization efforts to date target the position in the frequency spectrum and frequency width of elastic bandgaps~\cite{Sigmund2003, Hussein2006, Halkjaer2006, Gazonas2006, Bilal2011, Lu2017, Ronellenfitsch2019, Bacigalupo2019}. In this context, the techniques used are mainly gradient-based topology optimization and genetic algorithms~\cite{Li2019}. Fewer studies are dedicated to the optimization of other dispersive properties, e.g., the elastic wave speeds. Topology optimization has been used to design the transient response of one-dimensional elastic waveguides, based on a sensitivity analysis that uses the adjoint method to calculate the effects of topological changes on the group velocity ~\cite{Dahl2008}. A similar technique, albeit applied to frequency-domain equations, is adopted for spatial wave manipulation~\cite{Park2015, Andreassen2016, Christiansen2019}. In one example, Andreassen et al.\ design plates with perforations, optimized to control the direction of flexural modes at different frequencies~\cite{Andreassen2016}. However, in~\cite{Andreassen2016} the method avoids the calculation of implicit derivatives that describe the evolution of a mode shape with the cell architecture, an aspect that could be useful for the optimization of mode shape-dependent dispersion characteristics. Furthermore, the studies mentioned are restricted to structures consisting of a single material and void. However, we note that the local addition of a second material with a relatively high density can have drastic effects on the range of wave speeds that can be attained, while keeping the static stiffness constant\cite{matlack2016composite}.

In this article, we develop a tool to optimize the elastic wave speeds in three-dimensional architected lattices at a fixed frequency of propagation. The key step is a sensitivity analysis to calculate the topological gradient of the elastic group velocity of any particular mode at a chosen frequency of interest, by providing a method to compute the sensitivities of the mode shape together with an adjoint method. These derivatives are also useful in understanding and engineering the rich modal composition exhibited by these 3-D structures at any frequency. We show that analyzing the real and imaginary parts of the mode shape separately simplifies the problem of computing sensitivity. In order to demonstrate our method, we study five periodic lattices (simple cubic (SC), body-centered cubic (BCC), face-centered cubic (FCC), octet and hexagonal) that propagate pressure-like elastic waves at similar frequencies. We model these lattices using frame elements~\cite{liu2013finite, Hughes2012}. We use our sensitivity analysis to optimize the group velocity for a chosen mode and at a chosen frequency for all five structures, as a function of their density and/or static stiffness. The structural parameters we choose to vary are the circular cross-sectional areas of the beams and the value of point masses located at the joints. We non-dimensionalize our analysis, so that the results can be applied to any frequency, by simply scaling a length parameter within each unit cell. In order to verify our results, we compare them to finite element simulations with 3-D elements in COMSOL, and validate them through comparisons with experiments on 3-D printed samples. Our technique could also be applicable to other phenomena involving directional or preferential wave propagation, such as wave focusing or backscattering-free waveguiding.

Following this introduction, in Section~\ref{math}, we report details on the formulation of group velocity culminating with its sensitivity analysis. In Section~\ref{num}, we use our sensitivity analysis to calculate bounds on the wave speeds for the five considered lattices as a function of their densities and stiffness. Finally, in Section~\ref{exp}, we conduct experiments on two representative BCC lattices with engineered wave speeds and compare the results to our findings obtained from numerical calculations.

\section{Optimal design of group velocity} \label{math}
Several important applications, such as impedance-matched structures, delay lines, frequency filters and vibration absorbers require engineering wave speeds, i.e., group velocities, of specific modes at a frequency of propagation, or a range of frequencies. In this section, we define the group velocity of a wave propagating at a certain frequency in periodic media and we detail a framework to optimize the group velocity in a structure together with static properties such as the stiffness and mass density.

%%%%%%%%%%%%%%%%%%%%%%%%%%%%%%%%%%
%%%%%%%%%%%%%%%%%%%%%%%%%%%%%%%%%%
\subsection{Background}
Consider an infinite heterogeneous periodic medium, where the unit cell geometry is described by $p$ parameters $\underaccent{\tilde}\chi \in {\mathbb R}^p$.  In this work, we consider a discrete system (either a discrete mass system or one obtained by the discretization of a continuum system as in Section \ref{sec:frame}) with $d$ degrees of freedom $ \underaccent{\tilde}U \in {\mathbb R}^d$ in the reference unit cell.  We index the unit cells with $n$-tuple of integers $\underaccent{\tilde}n \in {\mathbb Z}^n$ depending on whether the unit cell is repeated in $n=1,2 \mbox{ or }3$ dimensions.   Let $u_{i\underaccent{\tilde}n}$ denote the generalized displacement and $p_{i\underaccent{\tilde}n} = \sum_{j,\underaccent{\tilde}m} M_{i\underaccent{\tilde}nj\underaccent{\tilde}m} \dot{u}_{j\underaccent{\tilde}m}$ denote the momentum associated with the $i^{th}$ degree of freedom in the $\underaccent{\tilde}n^{th}$ unit cell, and let $K_{i\underaccent{\tilde}nj\underaccent{\tilde}m}$ the stiffness between the  $i^{th}$ degree of freedom in the $\underaccent{\tilde}n^{th}$ unit cell and the  $j^{th}$ degree of freedom in the $\underaccent{\tilde}m^{th}$ unit cell.  Then the generalized equation of motion  is
\begin{equation} \label{eq:motion}
 \sum_{j,\underaccent{\tilde}m} M_{i\underaccent{\tilde}nj\underaccent{\tilde}m} \ddot{u}_{j\underaccent{\tilde}m} = F_{i\underaccent{\tilde}n} = \sum_{j, \underaccent{\tilde}m} K_{i\underaccent{\tilde}nj\underaccent{\tilde}m} (u_{i\underaccent{\tilde}n} - u_{j\underaccent{\tilde}m}) \quad i = 1, \dots d, \ \underaccent{\tilde}n \in {\mathbb Z}^n.
\end{equation}
The general solution to this equation may be written as a superposition of functions of the form $u_{i\underaccent{\tilde}n} (t) = \exp(-i\omega t) U_{i\underaccent{\tilde}n}$.  Further, Bloch's theorem ~\cite{Brillouin1953} enables us to write $U_{i\underaccent{\tilde}n} = \exp(i \underaccent{\tilde}k \cdot \underaccent{\tilde}n) U_i$ where $\underaccent{\tilde}U \in {\mathbb R}^d$ represents amplitude of displacements within a single unit cell.  
Thus, the general solution is a superposition of waves  $u_{i\underaccent{\tilde}n} (t) = \exp(i(\underaccent{\tilde}k \cdot \underaccent{\tilde}n - \omega t)) U_i$ with temporal frequency $\omega$ and wave vector $\underaccent{\tilde}{k}$.  Substituting this in eq. (\ref{eq:motion}), we obtain the characteristic equation
%Waves propagating through the medium, with wave vector $\underaccent{\tilde}k$ can be studied by solving the complex eigenvalue problem of the form \cite{bathe2006, Hughes2012}
\begin{equation}\label{eigenvalprob}
     \underaccent{\tilde}K(\underaccent{\tilde}\chi,\underaccent{\tilde}k)\underaccent{\tilde}U=\omega^2 \underaccent{\tilde}M(\underaccent{\tilde}\chi,\underaccent{\tilde}k)  \underaccent{\tilde}U,
\end{equation}
where $\underaccent{\tilde}K$ and $\underaccent{\tilde}M$ are the $d\times d$ self-adjoint stiffness and mass matrices with components
\begin{eqnarray}
K_{ij} (\underaccent{\tilde}\chi,\underaccent{\tilde}k) &=& \sum_{\underaccent{\tilde}n} \left(\exp(i \underaccent{\tilde}k \cdot \underaccent{\tilde}n) K_{i\underaccent{\tilde}0j\underaccent{\tilde}n}(\underaccent{\tilde}\chi) - \sum_{l} \delta_{ij} K_{j\underaccent{\tilde}0l\underaccent{\tilde}n}(\underaccent{\tilde}\chi)\right),\\
M_{ij} (\underaccent{\tilde}\chi,\underaccent{\tilde}k) &=&  \sum_{\underaccent{\tilde}n} \exp(i \underaccent{\tilde}k \cdot \underaccent{\tilde}n) M_{i\underaccent{\tilde}0j\underaccent{\tilde}n}(\underaccent{\tilde}\chi).
\end{eqnarray}

\begin{figure}
\centering
\includegraphics[width=.5\textwidth]{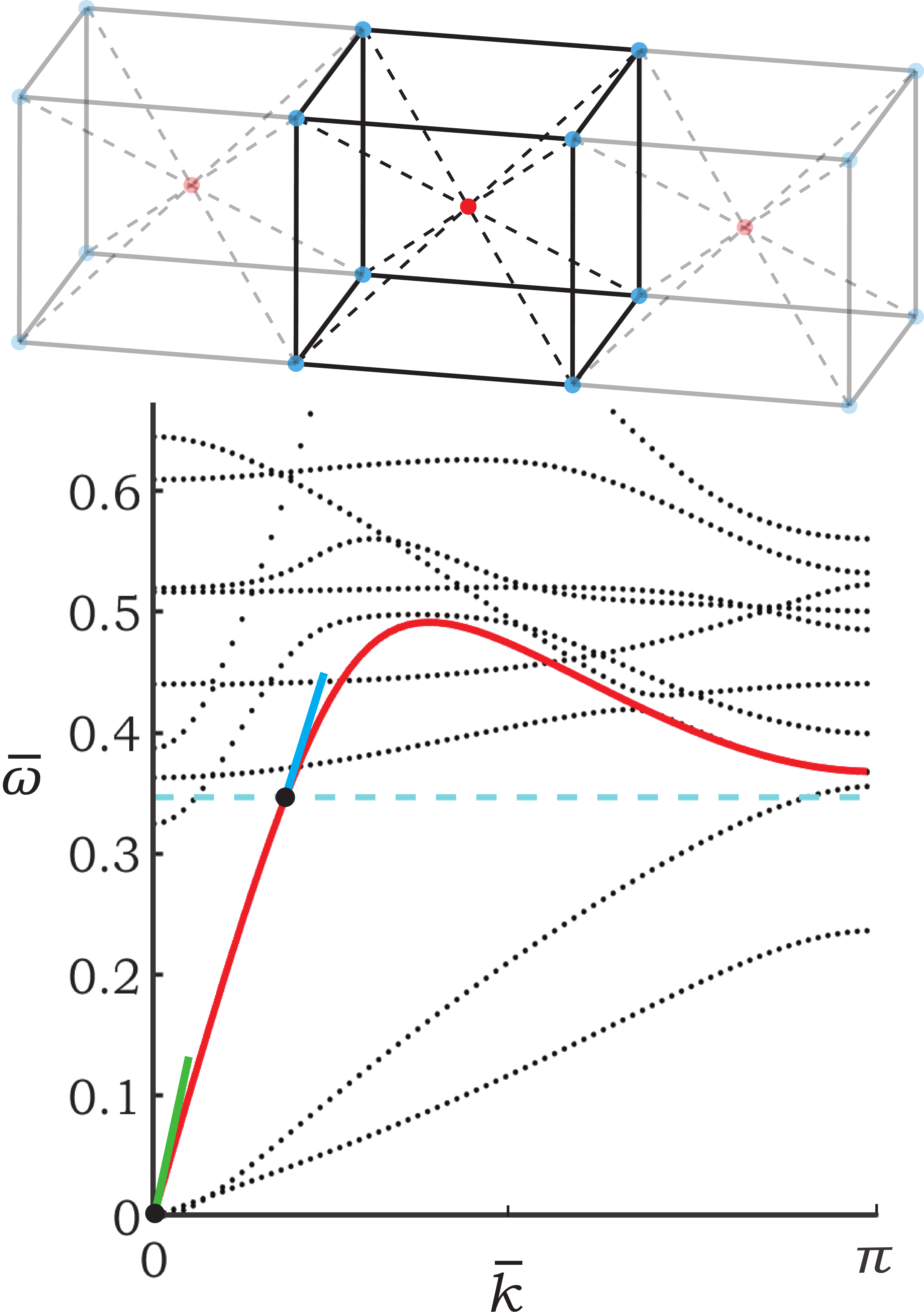}
\caption{\textit{ Example of a body-centered cubic lattice and the corresponding dispersion relation between non-dimensional frequency and wavenumber.}}
\label{fig:ex}
\end{figure}

Since $\underaccent{\tilde}n \in {\mathbb Z}^n$ is an $n$-tuple of integers and $\underaccent{\tilde}k$ occurs only in sinusoidal terms, it suffices to consider $\underaccent{\tilde}k$ in the irreducible Brillouin zone (${\mathcal B}$), which is the set $(0,\pi/L)^n$ for a cubic periodicity with period $L$.  Given any wave-vector $\underaccent{\tilde}k \in {\mathcal B}$, eq.~\eqref{eigenvalprob} is a generalized eigenvalue problem.  Further, since $\underaccent{\tilde}K, \underaccent{\tilde}M$ are self-adjoint and also positive definite for stable structures, there is a complete set of positive eigenvalues $\omega^2$. 
It is customary to normalize the eigenmodes $\underaccent{\tilde}U \cdot \underaccent{\tilde} M \underaccent{\tilde}U = 1$.
Since $\underaccent{\tilde}K, \underaccent{\tilde}M$ depend smoothly on $\underaccent{\tilde}k$, the resulting eigenvalues $\omega^2$ and eigenvectors (eigenmodes) $\underaccent{\tilde}U$ also depend smoothly on $\underaccent{\tilde}k$.
Thus, we obtain the \emph{dispersion relation} with $d$ branches $\omega^{(m)}(\underaccent{\tilde}k)$, $m = 1, \dots, d$.   A typical example -- that of a body-centered cubic lattice that repeats in one dimension is shown in  Fig. \ref{fig:ex}. The lattice displays point masses indicated by the red and blue dots at its joints. The plot shows the relation between non-dimensional frequency and non-dimensional wavenumber (magnitude of wave vector). We explain the non-dimensionalisation later in this paper. It is customary to identify branches with the form of the eigenmode $\underaccent{\tilde}U$ which may lead to crossings. For example, the branch associated with the longitudinal mode where all deformations are parallel to the direction of periodicity is indicated in red in Fig. \ref{fig:ex}.

In this work, we shall be concerned with wave packets consisting of waves with a particular mode with a narrow range of wave numbers and frequency.  The energy of a wave packet associated with the $m^{th}$ mode propagates with the \emph{group velocity} 
\begin{equation} \label{groupvela}
\underaccent{\tilde}C^{(m)} := {\partial \omega^{(m)} \over \partial \underaccent{\tilde}k} = \omega^{(m)}_{\underaccent{\tilde}k}.
\end{equation}
For future use we note the relations
\begin{equation} \label{groupvel}
    \underaccent{\tilde}C^{(m)}=\frac{1}{2\omega^{(m)}} \underaccent{\tilde}U^{(m)}.\left( \underaccent{\tilde}K_{\underaccent{\tilde}k}- (\omega^{(m)})^2 \underaccent{\tilde}M_{\underaccent{\tilde}k}\right)\underaccent{\tilde}U^{(m)}
\end{equation}
that follows from eq. (\ref{eigenvalprob}) and the normalization of the eigenmode. In the above, subscript indicates partial derivative of a quantity. The wave speed is calculated as the magnitude of group velocity, $C^{(m)}=|\underaccent{\tilde}C^{(m)}|$.

The static stiffness $E^{(m)}$ of the structure deformed in a particular mode is related to the speed of propagation of long wavelengths:
\begin{equation}
E^{(m)} = { (C^{(m)}_0)^2 \over \rho}, \quad \mbox{where} \quad  C^{(m)}_0 = \lim_{\underaccent{\tilde}k \to \underaccent{\tilde}0} \ \ C^{(m)} (\underaccent{\tilde}k)
\end{equation}
and $\rho$ is the average density of the unit cell. The long wavelength (or low frequency) wave speed and the finite wavelength (or moderate frequency) wave speed for the longitudinal mode in the illustrated BCC lattice are indicated by the slopes of the green and blue solid lines in Fig. \ref{fig:ex}, respectively.

Finally, we note that the dispersion relation may be such that there are no waves propagating at certain frequencies.  We call \emph{bandgaps} the frequency intervals in which no waves propagate.

%%%%%%%%%%%%%%%%%%%%%%%%%%%%%%%%%%
%%%%%%%%%%%%%%%%%%%%%%%%%%%%%%%%%%
\subsection{Optimal design problem} \label{opt_des}

We seek to design a structure, i.e.,  identify parameters $\underaccent{\tilde}\chi$ in a design set ${\mathcal D}$, so that the structure has a given static stiffness $E^*$, a given density $\rho^*$ and a given wave speed $C^*$ for a particular mode $m$ of propagation with a given frequency $\omega^*$.

The first step is to verify that there are indeed waves associated with mode $m$ and the given frequency $\omega^*$, i.e., we can find $\underaccent{\tilde} \chi_0 \in {\mathcal D}$ and $\underaccent{\tilde}k_0 \in {\mathcal B}$ such that $\omega^{(m)} (\underaccent{\tilde}\chi_0, \underaccent{\tilde}k_0) = \omega^*$.   Suppose we can identify such a point, and further suppose that the group velocity at that point is non-zero, i.e., $C^{(m)} (\underaccent{\tilde}\chi_0, \underaccent{\tilde}k_0) \ne 0$, then we can use the implicit function theorem to identify a function $\underaccent{\tilde}k =  \underaccent{\tilde} k^* (\underaccent{\tilde}\chi)$ such that
\begin{equation}
\omega^{(m)} (\underaccent{\tilde} \chi, \underaccent{\tilde}k^* (\underaccent{\tilde} \chi)) = \omega^*
\end{equation}
in the neighborhood of $\underaccent{\tilde}\chi_0, \underaccent{\tilde}k_0$.

Second, we consider the density.  The density $\rho(\underaccent{\tilde}\chi)$ is generally a monotone function of the parameters $\underaccent{\tilde}\chi$.  So it is generally possible to identify a set of admissible parameters which attain the given density:
\begin{equation}
{\mathcal D}_{\rho^*} = \{ \underaccent{\tilde}\chi \in {\mathcal D}: \rho(\underaccent{\tilde}\chi) = \rho^*\}.
\end{equation}
We assume that this is the case.  If not, we can proceed by adding a constraint on our optimal design problem below.

The third step is to understand the range of wave speeds that can be attained by the admissible set of parameters.  We do so by identifying the maximum and minimum wave speeds in terms of the optimal design problem
\begin{equation} \label{eq:optc}
 \underset{\underaccent{\tilde}\chi \in {\mathcal D}_{\rho^*}} {\text{max/ min}} \quad
C^{(m)} \left(\underaccent{\tilde}\chi,\underaccent{\tilde}k^* (\underaccent{\tilde}\chi) \right).
\end{equation}
We solve this problem using a descent method, but postpone the computation of the gradient till the next section.
%Figure \ref{fig:bounds}(a) shows a typical result that we shall discuss later.
%We then turn to bounding the set of wave speeds and elastic moduli pairs that one can attain using the admissible set of parameters.  
We derive a bound for the set of wave speeds and elastic moduli that one can attain using the admissible set of parameters, by studying the Pareto optimal of weighted averages. Given any $\gamma_1, \gamma_2 \in {\mathbb R}$, let
\begin{equation} \label{eq:optce}
 \mathcal{O} (\gamma_1, \gamma_2) := \underset{\underaccent{\tilde}\chi \in {\mathcal D}} {\text{max}} \quad
\gamma_1 C^{(m)} \left(\underaccent{\tilde}\chi,\underaccent{\tilde}k^* (\underaccent{\tilde}\chi) \right) + \gamma_2 E^{(m)} (\underaccent{\tilde} \chi).
\end{equation}
The convex hull of the set of all attainable wave speeds and stiffness is given by
\begin{equation}
{\mathcal S} = \{ (C^{(m)},E^{(m)}): \gamma_1 C^{(m)} + \gamma_2 E^{(m)} \le  \mathcal{O} (\gamma_1, \gamma_2), \gamma_1, \gamma_2 \in {\mathbb R} \}.
\end{equation}
We again solve (\ref{eq:optce}) using a descent method.
%Figure \ref{fig:bounds}(c) shows a typical result. 
We emphasize that this method only identifies the convex hull of the set: we do not know if the set is convex and therefore do not know how good an approximation the convex hull is to the actual set.  However, the method identifies extremal properties, which are of particular interest in the design of materials.

Finally, we address the full problem of identifying a structure that has a given static stiffness $E^*$, a given density $\rho^*$ and a given wave speed $C^*$ for a particular mode $m$ of propagation, with a given frequency $\omega^*$.  We can again pose this as an optimal design problem:  given $\delta_1, \delta_2 \ge 0 $,
\begin{equation} \label{eq:opt}
\underset{\underaccent{\tilde}\chi \in {\mathcal D}_{\rho^*}}{\text{min}} \ 
\left(\delta_1 \left( C^{(m)} \left(\underaccent{\tilde}\chi, \underaccent{\tilde}k^* \left( \underaccent{\tilde}\chi \right) \right) - C^* \right)^2 + \delta_2 \left( E^{(m)} \left(\underaccent{\tilde}\chi \right) - E^* \right)^2
 \right).
\end{equation}
This objective is non-negative and equal to zero exactly when we find an optimal design.  If the objective can not be driven to zero, then we have a design that does not meet all the requirements.  If this is the case, the ratio between $\delta_1$ and $\delta_2$ can be adjusted to find a design that matches the group velocity or the static stiffness.  Again, we solve (\ref{eq:opt}) using the a descent method, which requires the computation of the gradients.

\subsection{Sensitivity analysis} \label{sens_secn}
We address the optimization problems (\ref{eq:optc}), (\ref{eq:optce}) and (\ref{eq:opt}) using a descent method.   We start by calculating the sensitivity of the group velocity to the design variables.   Using the chain rule, we have:
\begin{equation}\label{sensa}
    \frac{d }{d \chi_j} \underaccent{\tilde}C^{(m)}\left( \underaccent{\tilde}\chi, \underaccent{\tilde}k^*(\underaccent{\tilde} \chi) \right)= 
    \frac{\partial \underaccent{\tilde}C^{(m)}}{\partial \chi_j} \left( \underaccent{\tilde}\chi, \underaccent{\tilde}k^*(\underaccent{\tilde} \chi) \right) + 
    \sum_l  \frac{\partial \underaccent{\tilde}C^{(m)}}{\partial k_l}\left( \underaccent{\tilde}\chi, \underaccent{\tilde}k^*(\underaccent{\tilde} \chi) \right) \times
     \frac{d k^*_l}{d \chi_j}\left( \underaccent{\tilde}\chi \right).
\end{equation}
Recalling that $\underaccent{\tilde}k^*$ is defined implicitly via the relation  $\omega^{(m)}\left(\underaccent{\tilde}\chi,\underaccent{\tilde}k^*(\underaccent{\tilde}\chi)\right)=\omega^*$. Let the direction of propagation of the wave be perpendicular to the plane containing unit vectors $\hat{\underaccent{\tilde}b}$ and $\hat{\underaccent{\tilde}c}$. We incorporate the constraints and calculate the sensitivities using the adjoint method, by modifying eq. \eqref{groupvela} as
\begin{multline} \label{adjointa}
    \underaccent{\tilde}C^{(m)}\left( \underaccent{\tilde}\chi, \underaccent{\tilde}k^*(\underaccent{\tilde} \chi) \right)={\partial \omega^{(m)} \over \partial \underaccent{\tilde}k} \left( \underaccent{\tilde}\chi, \underaccent{\tilde}k^*(\underaccent{\tilde} \chi) \right) +  \underaccent{\tilde}\lambda^{(1)} \left( \textrm{det} \left(\underaccent{\tilde}K\left( \underaccent{\tilde}\chi, \underaccent{\tilde}k^*(\underaccent{\tilde} \chi) \right)- {\omega^*}^2\underaccent{\tilde}M\left( \underaccent{\tilde}\chi, \underaccent{\tilde}k^*(\underaccent{\tilde} \chi) \right)\right) \right) + \\ \underaccent{\tilde}\lambda^{(2)} \left( \underaccent{\tilde}k^*(\underaccent{\tilde}\chi).\hat{\underaccent{\tilde}b}  \right)+ \underaccent{\tilde}\lambda^{(3)} \left( \underaccent{\tilde}k^*(\underaccent{\tilde}\chi).\hat{\underaccent{\tilde}c}  \right),
\end{multline}

where det(.) denotes the determinant of a matrix. Notice that det$\left(\underaccent{\tilde}K- {\omega^*}^2\underaccent{\tilde}M\right)=0$, when the relation $\omega^{(m)}=\omega^*$ is satisfied and $\underaccent{\tilde}k^*.\hat{\underaccent{\tilde}b}=\underaccent{\tilde}k^*.\hat{\underaccent{\tilde}c}=0$ when the wave vector is along $\hat{\underaccent{\tilde}k}=\hat{\underaccent{\tilde}b} \times \hat{\underaccent{\tilde}c}$. This allows us to choose adjoint variables $\underaccent{\tilde}\lambda^{(1)}$, $\underaccent{\tilde}\lambda^{(2)}$ and $\underaccent{\tilde}\lambda^{(3)}$ such that we avoid the computation of the derivative, $\dfrac{d k^*_l}{d \chi_j}$ in eq. \eqref{sensa}. Substituting eq. \eqref{adjointa} in eq. \eqref{sensa}, we get

\begin{multline} \label{sensb}
    \frac{d \underaccent{\tilde}C^{(m)}_i}{d \chi_j} = \frac{\partial^2 \omega^{(m)}}{\partial k_i \partial \chi_j} + \lambda^{(1)}_i \textrm{tr}\left( \textrm{adj}\left( \underaccent{\tilde}K- {\omega^*}^2\underaccent{\tilde}M \right) \left( \underaccent{\tilde}K_{\chi_j}- {\omega^*}^2\underaccent{\tilde}M_{\chi_j} \right) \right) + \\
    \sum_l \left( \frac{\partial^2 \omega^{(m)}}{\partial k_i \partial k_l} + \lambda^{(1)}_i \textrm{tr}\left( \textrm{adj}\left( \underaccent{\tilde}K- {\omega^*}^2\underaccent{\tilde}M \right) \left( \underaccent{\tilde}K_{k_l}- {\omega^*}^2\underaccent{\tilde}M_{k_l} \right) \right)  + 2 \lambda^{(2)}_i \sum_q \delta_{ql} \hat{b}_q  + 2 \lambda^{(3)}_i \sum_q \delta_{ql} \hat{c}_q \right)  \times \dfrac{d k^*_l}{d \chi_j},
\end{multline}
where tr(.) and adj(.) are the trace and adjoint of a matrix, respectively \cite{shores2007}. 

Let $\hat{a}_l=\textrm{tr}\left( \textrm{adj}\left( \underaccent{\tilde}K- {\omega^*}^2\underaccent{\tilde}M \right) \left( \underaccent{\tilde}K_{k_l}- {\omega^*}^2\underaccent{\tilde}M_{k_l} \right) \right)$. We pick $\underaccent{\tilde}\lambda^{(1)}$, $\underaccent{\tilde}\lambda^{(2)}$ and $\underaccent{\tilde}\lambda^{(3)}$ such that \\ $\left(\dfrac{\partial^2 \omega^{(m)}}{\partial^2 \underaccent{\tilde}k} + \underaccent{\tilde}\lambda^{(1)} \otimes \hat{\underaccent{\tilde}a} +2 \underaccent{\tilde}\lambda^{(2)} \otimes \hat{\underaccent{\tilde}b} +2 \underaccent{\tilde}\lambda^{(3)} \otimes \hat{\underaccent{\tilde}c}\right)=0$. This gives us 
\begin{align} \label{adjointvar}
    & \underaccent{\tilde}\lambda^{(1)}= -\frac{\dfrac{\partial^2 \omega^{(m)}}{\partial^2 \underaccent{\tilde}k} \hat{\underaccent{\tilde}k}}{\hat{\underaccent{\tilde}a}.\hat{\underaccent{\tilde}k}},\\ 
    & \underaccent{\tilde}\lambda^{(2)}= -\frac{\dfrac{\partial^2 \omega^{(m)}}{\partial^2 \underaccent{\tilde}k} \hat{\underaccent{\tilde}b} + \underaccent{\tilde}\lambda^{(1)} \left( \hat{\underaccent{\tilde}a}. \hat{\underaccent{\tilde}b} \right)}{2 \hat{\underaccent{\tilde}b}.\hat{\underaccent{\tilde}b}},\nonumber \\
    & \underaccent{\tilde}\lambda^{(3)}= -\frac{\dfrac{\partial^2 \omega^{(m)}}{\partial^2 \underaccent{\tilde}k} \hat{\underaccent{\tilde}c} + \underaccent{\tilde}\lambda^{(1)} \left( \hat{\underaccent{\tilde}a}. \hat{\underaccent{\tilde}c} \right)}{2 \hat{\underaccent{\tilde}c}.\hat{\underaccent{\tilde}c}}. \nonumber
\end{align}
Substituting eq. \eqref{adjointvar} into eq. \eqref{sensb} gives us
\begin{equation} \label{sensc}
    \frac{d \underaccent{\tilde}C^{(m)}_i}{d \chi_j} = \frac{\partial^2 \omega^{(m)}}{\partial k_i \partial \chi_j} -
    \frac{\sum_l \dfrac{\partial^2 \omega^{(m)}}{\partial k_i \partial k_l} \hat{k}_l}{\hat{\underaccent{\tilde}a}.\hat{\underaccent{\tilde}k}}   \textrm{tr}\left( \textrm{adj}\left( \underaccent{\tilde}K- {\omega^*}^2\underaccent{\tilde}M \right) \left( \underaccent{\tilde}K_{\chi_j}- {\omega^*}^2\underaccent{\tilde}M_{\chi_j} \right) \right) .
\end{equation}

Let $\alpha=\chi_j$ or $k_l$. Then, $\dfrac{\partial^2 \omega^{(m)}}{\partial k_i \partial \alpha}$ can be calculated by differentiating eq.~\eqref{groupvel} as
\begin{multline} \label{sens_doub}
    \dfrac{\partial^2 \omega^{(m)}}{\partial k_i \partial \alpha}=\frac{1}{2\omega^{(m)}}     \bigg(    2\,\mathrm{Re} \left(\underaccent{\tilde}U^{(m)}_\alpha.\left(\underaccent{\tilde}K_{k_i} - {(\omega^{(m)})}^2 \underaccent{\tilde}M_{k_i}\right)\underaccent{\tilde}U^{(m)}\right)    + \\
    \underaccent{\tilde}U^{(m)}.\left(\underaccent{\tilde}K_{k_i\alpha}-{(\omega^{(m)})}^2 \underaccent{\tilde}M_{k_i\alpha} -2\omega^{(m)} \omega^{(m)}_\alpha \underaccent{\tilde}M_{k_i}\right)\underaccent{\tilde}U^{(m)} - 
    \frac{\omega^{(m)}_\alpha}{\omega^{(m)}}\underaccent{\tilde}U^{(m)}.\left(\underaccent{\tilde}K_{k_i} - {(\omega^{(m)})}^2 \underaccent{\tilde}M_{k_i}\right)\underaccent{\tilde}U^{(m)} \bigg),
\end{multline}
where $\mathrm{Re}(.)$ is the real part of a complex entity. 

Evaluating eq.~\eqref{sens_doub} poses a challenge, as the sensitivity of the eigenvector with respect to the topology and wavenumber, $\underaccent{\tilde}U^{(m)}_\alpha$, is not trivial to calculate. This is because the explicit expression of the mode shape $\underaccent{\tilde}U^{(m)}$ in terms of $\chi_j$ and $k_l$ is often difficult to compute. 

We now present a method to calculate the sensitivities $\underaccent{\tilde}U^{(m)}_\alpha$, which can be used to compute the sensitivities of dispersion properties that depend on the eigenvector $\underaccent{\tilde}U^{(m)}$. By analysing the real and imaginary parts of eq.~\eqref{eigenvalprob} separately, we arrive at the equations
\begin{align} 
    &\underaccent{\tilde}A^R {\underaccent{\tilde}U^{(m)}}^R - \underaccent{\tilde}A^I {\underaccent{\tilde}U^{(m)}}^I=0,\label{real_eigvalprob} \\
    &\underaccent{\tilde}A^I {\underaccent{\tilde}U^{(m)}}^R + \underaccent{\tilde}A^R {\underaccent{\tilde}U^{(m)}}^I=0, \label{img_eigvalprob}
\end{align}
where $\underaccent{\tilde}A= \underaccent{\tilde}K- {(\omega^{(m)})}^2 \underaccent{\tilde}M$ and the superscripts $R$ and $I$ denote the real and imaginary part of the complex entity. The mass orthogonality constraint $\underaccent{\tilde}U^{(m)}.\underaccent{\tilde}M \underaccent{\tilde}U^{(m)}=1$ can be written as 
\begin{equation} \label{massog}
\left( {\underaccent{\tilde}U^{(m)}}^R - i {\underaccent{\tilde}U^{(m)}}^I \right). \left( \left( \underaccent{\tilde}M^R {\underaccent{\tilde}U^{(m)}}^R -\underaccent{\tilde}M^I {\underaccent{\tilde}U^{(m)}}^I \right)  + i \left( \underaccent{\tilde}M^I {\underaccent{\tilde}U^{(m)}}^R +\underaccent{\tilde}M^R {\underaccent{\tilde}U^{(m)}}^I \right)  \right)=1,
\end{equation}
Differentiating equations \eqref{real_eigvalprob} to \eqref{massog} with respect to $\alpha$ yields
\begin{equation} 
    \underaccent{\tilde}A^R {\underaccent{\tilde}U^{(m)}_\alpha}^R - \underaccent{\tilde}A^I {\underaccent{\tilde}U^{(m)}_\alpha}^I=\underaccent{\tilde}A^I_\alpha {\underaccent{\tilde}U^{(m)}}^I - \underaccent{\tilde}A^R_\alpha {\underaccent{\tilde}U^{(m)}}^R=:b_1,\label{real_eigvalprob_diff} 
\end{equation}
\begin{equation}
    \underaccent{\tilde}A^R {\underaccent{\tilde}U^{(m)}_\alpha}^I + \underaccent{\tilde}A^I {\underaccent{\tilde}U^{(m)}_\alpha}^R=-\underaccent{\tilde}A^I_\alpha {\underaccent{\tilde}U^{(m)}}^R - \underaccent{\tilde}A^R_\alpha {\underaccent{\tilde}U^{(m)}}^I=:b_2,\label{img_eigvalprob_diff} 
\end{equation}
\begin{equation}
        {\underaccent{\tilde}U^{(m)}_\alpha}^R . \left(\underaccent{\tilde}M^R {\underaccent{\tilde}U^{(m)}}^R\right) + {\underaccent{\tilde}U^{(m)}_\alpha}^I . \left(\underaccent{\tilde}M^R {\underaccent{\tilde}U^{(m)}}^I\right) = -\frac{1}{2}\left({\underaccent{\tilde}U^{(m)}}^R. \underaccent{\tilde}M^R_\alpha {\underaccent{\tilde}U^{(m)}}^R + {\underaccent{\tilde}U^{(m)}}^I. \underaccent{\tilde}M^R_\alpha {\underaccent{\tilde}U^{(m)}}^I  \right)         =:c_1,\label{real_massog_diff} 
\end{equation}
\begin{equation}
        {\underaccent{\tilde}U^{(m)}_\alpha}^R . \left(\underaccent{\tilde}M^I {\underaccent{\tilde}U^{(m)}}^R\right) + {\underaccent{\tilde}U^{(m)}_\alpha}^I . \left(\underaccent{\tilde}M^I {\underaccent{\tilde}U^{(m)}}^I\right) = -\frac{1}{2}\left({\underaccent{\tilde}U^{(m)}}^R. \underaccent{\tilde}M^I_\alpha {\underaccent{\tilde}U^{(m)}}^R + {\underaccent{\tilde}U^{(m)}}^I. \underaccent{\tilde}M^I_\alpha {\underaccent{\tilde}U^{(m)}}^I  \right)         =:c_2.\label{img_massog_diff}
\end{equation}
We notice that equations \eqref{real_eigvalprob} to \eqref{img_eigvalprob_diff} together with the sum of equations \eqref{real_massog_diff} and \eqref{img_massog_diff} can be combined using a block matrix representation as
\begin{equation} \label{matrixeq}
    \begin{pmatrix} 
0 & \left(\left(\underaccent{\tilde}M^R+\underaccent{\tilde}M^I\right){\underaccent{\tilde}U^{(m)}}^R\right)^T & \left(\left(\underaccent{\tilde}M^R+\underaccent{\tilde}M^I\right){\underaccent{\tilde}U^{(m)}}^I\right)^T \\ \\
-b_1 & \underaccent{\tilde}A^R & -\underaccent{\tilde}A^I \\ \\
-b_2 & \underaccent{\tilde}A^I & \underaccent{\tilde}A^R 
\end{pmatrix}
\begin{pmatrix}
1 \\ \\
{\underaccent{\tilde}U^{(m)}_\alpha}^R-(c_1+c_2){\underaccent{\tilde}U^{(m)}}^R\\ \\
{\underaccent{\tilde}U^{(m)}_\alpha}^I-(c_1+c_2){\underaccent{\tilde}U^{(m)}}^I
\end{pmatrix}=
\begin{pmatrix}
0 \\ \\
0\\ \\
0
\end{pmatrix}.
\end{equation}
Superscript $T$ denotes transpose of a matrix. Now, set
\begin{align}
&D_1=\begin{pmatrix}
0 & \left(\left(\underaccent{\tilde}M^R+\underaccent{\tilde}M^I\right){\underaccent{\tilde}U^{(m)}}^R\right)^T & \left(\left(\underaccent{\tilde}M^R+\underaccent{\tilde}M^I\right){\underaccent{\tilde}U^{(m)}}^I\right)^T \\ \\
-b_1 & \underaccent{\tilde}A^R & -\underaccent{\tilde}A^I \\ \\
-b_2 & \underaccent{\tilde}A^I & \underaccent{\tilde}A^R 
\end{pmatrix},\ D_2=\begin{pmatrix} 
\underaccent{\tilde}A^R & -\underaccent{\tilde}A^I \\ \\
\underaccent{\tilde}A^I & \underaccent{\tilde}A^R 
\end{pmatrix}, \nonumber \\
&D_3=\begin{pmatrix} 
0 & 0 & 0 \\ \\
0 & \underaccent{\tilde}A^R & -\underaccent{\tilde}A^I \\ \\
0 & \underaccent{\tilde}A^I & \underaccent{\tilde}A^R 
\end{pmatrix},\ D_4=\begin{pmatrix} 
0 & \left(\left(\underaccent{\tilde}M^R+\underaccent{\tilde}M^I\right){\underaccent{\tilde}U^{(m)}}^R\right)^T & \left(\left(\underaccent{\tilde}M^R+\underaccent{\tilde}M^I\right){\underaccent{\tilde}U^{(m)}}^I\right)^T \\ \\
-b_1 & 0 & 0 \\ \\
-b_2 & 0 & 0 
\end{pmatrix}. 
\end{align}

Let the size of matrix $D_1$ be $n\times n$. We have that rank$(D_2)$ is $n-2$, assuming single multiplicity of the eigenvalue at the fixed frequency $\omega^*$ for a chosen mode in eq.~\eqref{eigenvalprob}. As $D_2$ is a submatrix of $D_1$, rank$(D_1)\geq n-2$. From the previous argument, it is straightforward to show that rank$(D_3)=n-3$ and we also see that $D_4$ is a matrix of rank at most equal to one. Since, $D_1$ is the sum of matrices $D_3$ and $D_4$, we have that rank$(D_1)\leq n-2$. Hence, we conclude that the nullity of matrix $D_1$ is two. Let the elements of this null space be represented by $\underaccent{\tilde}y_1$ and $\underaccent{\tilde}y_2$. Then, eq.~\eqref{matrixeq} gives us

\begin{equation} \label{dispsens}
\begin{pmatrix}
0 \\ \\
{\underaccent{\tilde}U^{(m)}_\alpha}^R\\ \\
{\underaccent{\tilde}U^{(m)}_\alpha}^I
\end{pmatrix}=
\beta_1 \underaccent{\tilde}y_1 + \beta_2 \underaccent{\tilde}y_2 - 
\begin{pmatrix}
1 \\ \\
-(c_1 + c_2){\underaccent{\tilde}U^{(m)}}^R\\ \\
-(c_1 + c_2){\underaccent{\tilde}U^{(m)}}^I
\end{pmatrix}.
\end{equation}
Finally, the sensitivities $\underaccent{\tilde}U^{(m)}_\alpha$ can be computed as $\underaccent{\tilde}U^{(m)}_\alpha={\underaccent{\tilde}U^{(m)}_\alpha}^R + i {\underaccent{\tilde}U^{(m)}_\alpha}^I$. The scalars $\beta_1$ and $\beta_2$ can be calculated from the equality that the first element of the vector $\beta_1 \underaccent{\tilde}y_1 + \beta_2 \underaccent{\tilde}y_2$ is 1, together with one of equations \eqref{real_massog_diff} or \eqref{img_massog_diff}. For cases where the multiplicity of the eigenvalue $\omega^*$ is more than one, we have multiple eigenvectors satisfying the eigenvalue problem given by eq.~\eqref{eigenvalprob} and as a result end up with more than two vectors in the null space from eq.~\eqref{matrixeq}. This requires additional information on the sensitivities of the desired mode shape. Such situations seldom happen and, for this study,  we select geometries and modes where the eigenvalue from eq.~\eqref{eigenvalprob} has single multiplicity. 

Plugging the sensitivities $\underaccent{\tilde}U^{(m)}_\alpha$ into eq.~\eqref{sens_doub} and subsequently into eq.~\eqref{sensc} gives us the sensitivities of the group velocity with respect to the topological parameters. The sensitivities of the objectives in problems (\ref{eq:optc}), (\ref{eq:optce}) and (\ref{eq:opt}) can be expressed in terms of the sensitivity of the group velocity, to solve them using a descent method. In the rest of the paper, we look at waves propagating along one of the orthogonal axes, which is the direction of periodicity of a lattice, i.e., $\hat{\underaccent{\tilde}k}=(1, 0, 0)^T, \hat{\underaccent{\tilde}b}=(0, 1, 0)^T, \hat{\underaccent{\tilde}c}=(0, 0, 1)^T$, and $\hat{\underaccent{\tilde}a}$ is parallel to $\hat{\underaccent{\tilde}k}$. As we study waves propagating in this fixed direction, any function of the wave vector can be thought of as a function of wavenumber in that direction. We refer to the component of the group velocity vector from eq. \eqref{groupvela} along the direction of propagation as the group velocity or wave speed.

%%%%%%%%%%%%%%%%%%%%%%%%%%%%%
%%%%%%%%%%%%%%%%%%%%%%%%%%%%%
\section{Lattices} \label{num}
\begin{figure}
\centering
\includegraphics[width=1.\textwidth]{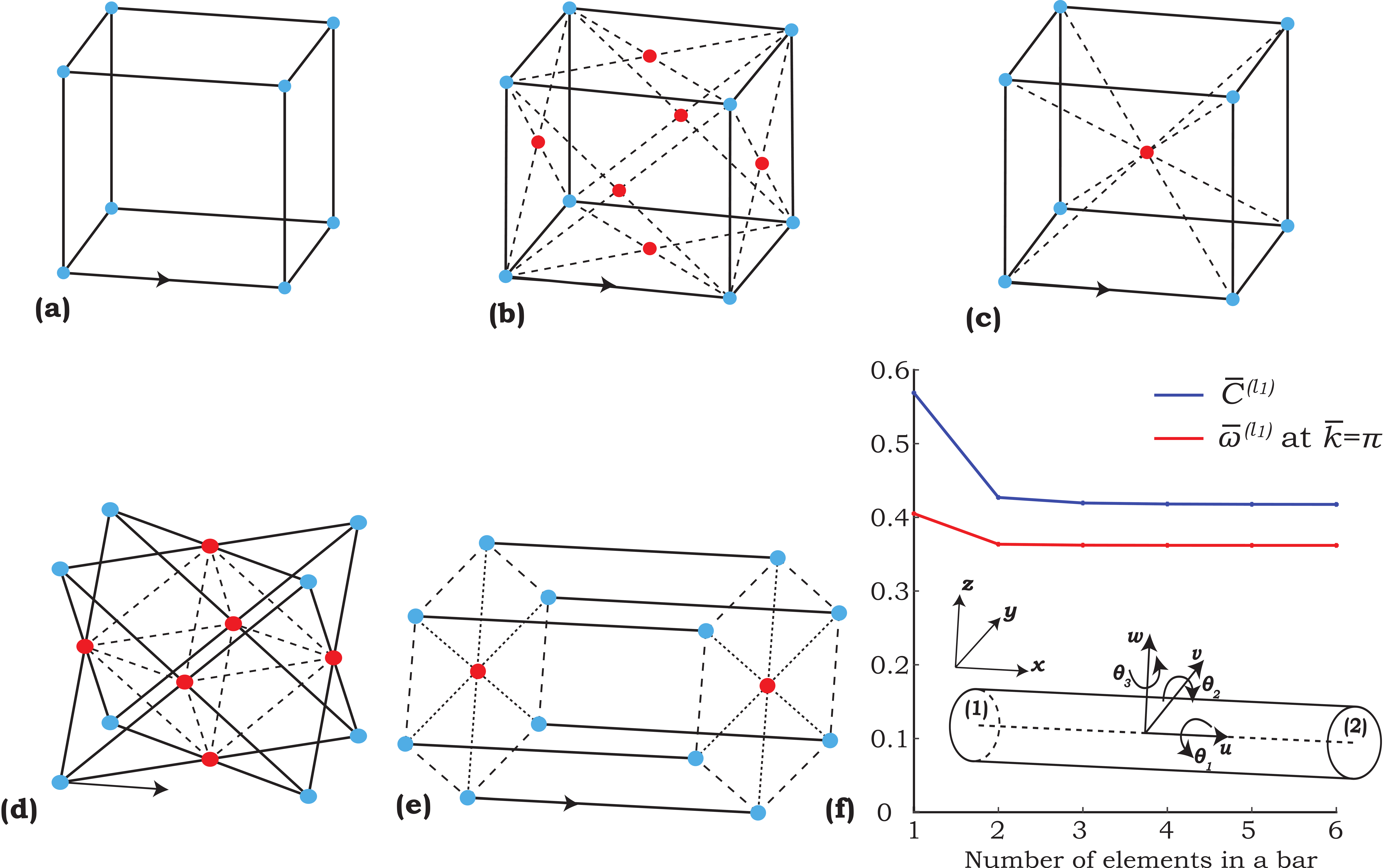}
\caption{\textit{\textbf{(a)}-\textbf{(e)} Unit cells from the SC, FCC, BCC, octet and hexagonal lattices, respectively, where the lines are links, the colored dots are concentrated, point masses and the arrows indicate the direction of wave propagation. \textbf{(f)}  Convergence study of wave speed and frequency of the first longitudinal mode with number of frame elements in a bar for a BCC lattice. Inset shows a frame element with two nodes and six degrees of freedom at each node.}} 
\label{fig_2}
\end{figure}

In this section, we solve the optimal design problems using the sensitivity analysis explained in the previous section to study five periodic lattice structures with masses at joints as shown in Fig. \ref{fig_2}(a)-(e). However, the methods developed in this paper are general and can be used on structures with arbitrary directions of periodicity and arbitrarily chosen frequencies of propagation.

We repeat each unit cell along one direction indicated by the arrow in Fig.~\ref{fig_2}(a)-(e), with a lattice parameter equal to the length of the unit cell in that direction, as to generate one dimensional arrays of cells. For these arrays, we consider longitudinal waves propagating at a fixed frequency along the direction indicated by the arrow, and solve the optimal design problems from section \ref{opt_des}. The topological parameters varied are the circular cross-sectional areas of the bars together with point masses at each node, while still maintaining symmetry of the structures. In each unit cell, bars indicated by the same stroke (solid, dashed or dotted) are constrained to have identical cross-sectional areas. Similarly, the points at the corner nodes (shown in blue) have the same mass in each unit cell and all the points marked in red have the same mass.

\subsection{Frame elements} \label{sec:frame}
 We discretize each bar in a lattice connecting two point masses, into multiple frame elements. A frame element can deform along its axis as well as along the directions perpendicular to its axis. Each element can carry axial and shear loads, and bending and torsional moments. Hence, each point on the element has six degrees of freedom- three translational displacements ($u$, $v$ and $w$) and three rotational ($\theta_1$, $\theta_2$ and $\theta_3$),  as shown in Fig. \ref{fig_2}(f). Assuming small deformation, these displacements and angles can be calculated as 
\begin{align}
    & u= N_1 u^{(1)} + N_2 u^{(2)}, \\ \nonumber
    & v= N_3 v^{(1)} + N_4 \theta_3^{(1)} +  N_5 v^{(2)} + N_6 \theta_3^{(2)}, \\ \nonumber
    & w= N_3 w^{(1)} + N_4 \theta_2^{(1)} +  N_5 w^{(2)} + N_{6} \theta_2^{(2)}, \\ \nonumber
    & \theta_1= N_{1} \theta_1^{(1)} + N_{2} \theta_1^{(2)}, \\ \nonumber
    & \theta_2= \frac{\partial w}{\partial x}, \\ \nonumber
    & \theta_3= \frac{\partial v}{\partial x}. \nonumber
\end{align}
Here, the degrees of freedom $u^{(1)},\ v^{(1)},\ w^{(1)},\ \theta^{(1)}_1,\ \theta^{(1)}_2$ and $\theta^{(1)}_3$ (three translational (displacement) degrees of freedom and three rotational (angular) degrees of freedom) correspond to one end of a frame element (labeled (1) in Fig. \ref{fig_2}(f)). The degrees of freedom $u^{(2)},\ v^{(2)},\ w^{(2)},\ \theta^{(2)}_1,\ \theta^{(2)}_2$ and $\theta^{(2)}_3$ are the corresponding degrees of freedom associated with the other end (labeled (2)) of the frame element. The shape functions $N_i,\ i=1,\dots, 6$ can be calculated from the boundary conditions as \cite{Hughes2012,liu2013finite}
\begin{align}
    & N_1=\frac{1-\xi}{2}, \\ \nonumber
    & N_2=\frac{1+\xi}{2}, \\ \nonumber
    & N_3=\frac{2-3\xi+\xi^3}{4}, \\ \nonumber
    & N_4=\frac{a(1-\xi-\xi^2+\xi^3)}{4}, \\ \nonumber
    & N_5=\frac{2+3\xi-\xi^3}{4}, \\ \nonumber
    & N_6=\frac{a(-1-\xi+\xi^2+\xi^3)}{4}, \nonumber
\end{align}
where $a=L_{el}/2$, with $L_{el}$ the length of the element, and $\xi=x/a$, with $-a \leq x \leq a$ and $x=0$ and the center of the element.

From this point onward, we non-dimensionalise our analysis so that the designs derived in the following sections can be scaled to any frequency of operation with just a length parameter of the lattice. Let $L$ be the length of a unit cell along the direction of periodicity, $\rho_s$ be the density of the solid material and $E_s$ be the Young's modulus of elasticity of the solid material. Then, every length, mass and time dimension is non-dimensionalised by dividing them with $L$, $\rho_s L^3$ and $\sqrt{\frac{\rho_s}{E_s}}L$, respectively.

We non-dimensionalize the elemental stiffness and mass matrices for frame elements from ref. \cite{Hughes2012,liu2013finite}, as detailed in Appendix A. Assembling them for a wave propagating through the structure, while invoking Bloch's theorem gives us the non-dimensional version of eq. \eqref{eigenvalprob}

\begin{equation} \label{globaleq}
         \overline{\underaccent{\tilde}K}\left(\overline{\underaccent{\tilde}\chi},\overline{\underaccent{\tilde}k}\right) \overline{\underaccent{\tilde}U}\left(\overline{\underaccent{\tilde}\chi},\overline{\underaccent{\tilde}k}\right)=\overline{\omega}\left(\overline{\underaccent{\tilde}\chi},\overline{\underaccent{\tilde}k}\right)^2\  \overline{\underaccent{\tilde}M}\left(\overline{\underaccent{\tilde}\chi},\overline{\underaccent{\tilde}k}\right)  \overline{\underaccent{\tilde}U}\left(\overline{\underaccent{\tilde}\chi},\overline{\underaccent{\tilde}k}\right),
\end{equation}
where $\overline{\underaccent{\tilde}K}$ and $\overline{\underaccent{\tilde}M}$ are the non-dimensional self-adjoint global stiffness and mass matrices, and $\overline{\underaccent{\tilde}U}$ is the non-dimensional global displacement vector. $\overline{\omega}$ and $\overline{\underaccent{\tilde}k}$ represent the non-dimensional frequency and wave vector of propagation. The vector $\overline{\underaccent{\tilde}\chi}$ represents the topology of a unit cell and comprises the non-dimensional cross sectional areas of bars (cross-sectional area divided by $L^2$) and the non-dimensional masses at the joints (mass divided by $\rho_s L^3$). Note that for a given frequency of wave propagation $\omega^*$ (in rad/s), the corresponding non-dimensional frequency $\overline{\omega}^*$ (in rad) can be calculated as $\omega^* \sqrt{\frac{\rho_s}{E_s}} L$. Hence, the optimal topology obtained from optimizing the wave speed at one particular frequency can be scaled by a function of the length of the unit cell $L$, in order to obtain the topology that optimizes the wave speed at any chosen frequency. 

We perform a convergence study to test the effect of variations in the  number of elements considered in a bar on the lattices' dispersion properties, as shown in Fig. \ref{fig_2}(f). As an example, we consider a BCC lattice without any point masses at the joints, and edge bars (solid in Fig. \ref{fig_2}(c)) four times in cross-sectional area compared to the diagonal bars (dotted). We first measure the non-dimensional wave speed of the first longitudinal mode at a non-dimensional frequency $\overline{\omega}^*=0.35$ (illustrated by the slope of the blue solid line in Fig. \ref{fig:ex}). Second, we measure the non-dimensional frequency of the mode (red curve in Fig. \ref{fig:ex}) at a high non-dimensional wavenumber $\overline{k}=|\overline{\underaccent{\tilde}k}|=\pi$, which tends to be sensitive to the number of elements considered. The change in wave speed is $1.70\%$ and in frequency at $\overline{k}=\pi$ is $0.33\%$, on increasing the number of elements in a bar from two to three. As a result of this negligible increase, we consider in the rest of this paper, two elements in each bar.

In the following subsection, we calculate bounds on the non-dimensional group velocity of an $n^{th}$ longitudinal mode, $\overline{C}^{(l_n)}$ as a function of non-dimensional density and longitudinal static stiffness of each 1-D periodic lattice, for a non-dimensional frequency of propagation $\overline{\omega}^*=0.35$. To compute these bounds, we solve the problems formulated in section \ref{opt_des}, which are non-dimensionalized as described earlier in this section.

\subsection{Bounds on group velocity} \label{sec:numbounds}
For waves propagating along the directions indicated by the arrows in Fig.~\ref{fig_2}(a)-(e), the non-dimensional wavenumber $\overline{k}$ (in rad) ranges from $0$ to $\pi$ on $\mathcal{B}$. This zone corresponds to the path that the wave vectors are restricted to in section \ref{math}. We can calculate the non-dimensional frequency $\overline{\omega}$ associated with each $\overline{k}$ for a fixed topology of a unit cell, to obtain its band structure. An example of this dispersion relation is shown in Fig. \ref{fig:ex}, for a BCC lattice. Each frequency of propagation excites multiple modes within the lattice, highlighting the modal richness typical of lattice structures. The mode highlighted in red indicates the first longitudinal mode of the structure, i.e., the frequencies of this mode at any wavenumber correspond to $\overline{\omega}^{(l_1)}$. This longitudinal mode is the combination of a mode that involves displacements only along the direction of propagation of the wave and a flexural mode that involves bending of the beams along this direction. We obtain this longitudinal mode by isolating its frequency of propagation from the other modes. The algorithm used to obtain the frequency of propagation $\overline{\omega}^{(l_n)}$ of the  $n^{th}$ longitudinal mode at any given wavenumber $\overline{k}$ is discussed in Algorithm~\ref{alg:alg1} of Appendix~B. 

\begin{table}  
  \begin{center}
    \caption{Material properties and bounds on topological variables}
   \label{params}
    \begin{tabular}{c||c|c|c|c|c|c|c} 
      \textbf{Parameter} & $E_s$ & $\rho_s$ & $G_s$ & $\overline{A}_{min}$ & $\overline{A}_{max}$ & $\overline{m}_{min}$ & $\overline{m}_{max}$  \\
      \hline
      \rule{0pt}{3ex}
      \textbf{Value} & 1.7 GPa & 930 kg/m$^3$ & 0.63 GPa & $4.25 \times 10^{-3}$ & $17 \times 10^{-3}$ & 0 & $34 \times 10^{-3}$ 
    \end{tabular}
  \end{center}
\end{table}

First, we calculate bounds on the non-dimensional group velocity as a function of density, for each of the five lattices repeating in one direction, as shown in Fig. \ref{fig_2} at $\overline{\omega}^*= 0.35$, by solving problem \eqref{eq:optc}. %Note that, in a dispersion relation as the one in Fig.~\ref{fig_2}(g), the group velocity is visually represented by the local slope of a certain branch, at the frequency of operation $\overline{\omega}_0$ indicated by the blue dashed line. 
%To avoid the presence of bandgaps at the chosen frequency $\overline{\omega}^*$,
We optimize the wave speed of the first longitudinal mode in the SC, BCC, hexagonal and octet lattices and third longitudinal mode in the FCC lattice, as these are the longitudinal modes propagated in each lattice at frequency $\overline{\omega}^*$, within the bounds of design variables discussed below.

We assign material properties of the solid material (Table \ref{params}) making the lattice, to be that of Polyamide 12~\cite{Amado2008, Pajunen2019}, which we also use for our experimental validation. $G_s$ is the shear modulus of rigidity of the material. We set constraints on the topology of each lattice (domain $\mathcal{D}$ in section~\ref{opt_des}), by choosing bounds on the non-dimensional cross-sectional areas of the bars and on the added masses, concentrated at the joints. In our analysis, the minimum and maximum non-dimensional areas of a bar are $\overline{A}_{min}$ and $\overline{A}_{max}$, respectively, and minimum and maximum mass at a joint are $\overline{m}_{min}$ and $\overline{m}_{max}$, respectively, indicated in Table \ref{params}. We pick these bounds as an example where bars remain slender, but the bounds can be chosen based on the manufacturing tolerances for specific applications.
%We pick these bounds so that the beams are slender enough to be modeled by our numerical method, while still ensuring their structural integrity to repeatedly test the 3-D printed samples. The maximum mass ($\overline{m}_{max}$) at a joint is considered to be twice as heavy as the thickest possible beam, with a length equal to the lattice parameter. The minimum mass ($\overline{m}_{min}$) is taken to be zero. 
We can represent these bounds on the topology vector $\overline{\underaccent{\tilde}\chi}$ by non-dimensional lower and upper bound vectors $\underaccent{\tilde}B_l$ and $\underaccent{\tilde}B_u$, respectively. As the topology of each unit cell is described by a vector of cross-sectional areas and masses, it is easy to calculate the non-dimensional density of a lattice from a linear equation of the form $f\left( \overline{\underaccent{\tilde}\chi} \right)=\underaccent{\tilde}A_{eq}. \overline{\underaccent{\tilde}\chi}= \overline{\rho}$. We solve the optimization problems with MATLAB's gradient based solver \textit{fmincon}, which also allows for linear equality and inequality constraints.

To determine an initial guess for topology and wavenumber and to ensure we start with a geometry that has wave propagation at $\overline{\omega}^*$, we solve the problem,
\begin{equation}\label{initg_nondim}
\begin{aligned}
& \underset{\overline{\underaccent{\tilde}\chi},\ \overline{k}}{\text{arg min}}
& & \left(\overline{\omega}^*- \overline{\omega}^{(l_n)}\left( \overline{\underaccent{\tilde}\chi}, \overline{k} \right) \right)^2 \\
& \text{subject to}
& & 0 \leq \overline{k} \leq \pi \\
&&& \underaccent{\tilde}B_l \leq \overline{\underaccent{\tilde}\chi} \leq \underaccent{\tilde}B_u \\
&&& \underaccent{\tilde}A_{eq}. \overline{\underaccent{\tilde}\chi}= \overline{\rho},\end{aligned}
\end{equation}
using Algorithm \ref{alg:alg1} together with sensitivities derived in section~\ref{sens_secn}, using a gradient based optimization method. Given a topology of a structure $\overline{\underaccent{\tilde}\chi}_0$ with wave propagation of the desired longitudinal mode at $\overline{\omega}^*$, the corresponding wavenumber of propagation $\overline{k}_0\left( \overline{\underaccent{\tilde}\chi}_0 \right)$ can be calculated by solving 
\begin{equation}\label{freqtownum}
\begin{aligned}
& \underset{\overline{k}}{\text{arg min}}
& & \left(\overline{\omega}^*- \overline{\omega}^{(l_n)}\left( \overline{\underaccent{\tilde}\chi}_0, \overline{k} \right) \right)^2 \\
& \text{subject to}
& & 0 \leq \overline{k} \leq \pi. \end{aligned}
\end{equation}

We now solve problem~\eqref{eq:optc}, to calculate bounds on the non-dimensional group velocity at a given non-dimensional density as follows. Starting from the initial guess for topology from problem~\eqref{initg_nondim}, the gradient of the wave speed with respect to the unit cell topology, from section \ref{sens_secn} is used to increment the topology vector along the direction of steepest descent/ ascent (for minimization/ maximization). After this increment, the new wavenumber corresponding to the updated topology $\overline{k} \left( \overline{\underaccent{\tilde}\chi} \right)$ at frequency $\overline{\omega}^*$ is calculated by solving problem~\eqref{freqtownum}. These increments are made until we converge to an optimum around the initial guess. Owing to the non-linearity of the problem, the topologies corresponding to the optimal values need not be unique.

For some geometries, we notice that more than one wavenumber can coexist at a given frequency for a selected mode, due to the fact that dispersion branches need not be monotonic. For example, this is the case for the highlighted longitudinal mode in Fig. \ref{fig:ex}. In such cases, we optimize the wave speed that corresponds to a positive group velocity. As a last note on the optimization problem, we fix the length of the diagonal bars in the hexagonal lattice as half the length of the unit cell along the direction of periodicity, so that we maintain the linear form of the function describing its density in terms of $\overline{\underaccent{\tilde}\chi}$.

%We use these values of $\overline{\omega}^{(L_n)}$ together with the sensitivities derived in the previous section to solve a non-dimensional version of problem~\eqref{initg} using a gradient based approach to determine the initial guess for topology. Starting from this initial guess, to solve the non-dimensional versions of problems~\eqref{mainoptprob}-\eqref{mainoptprob_stiff}, \eqref{mainoptprob_mod} and \eqref{mainoptprob_stiff_mod}, we can obtain the wavenumber of propagation, i.e. $|\overline{\underaccent{\tilde}k} \left( \overline{\underaccent{\tilde}\chi} \right)|$  of the $n^{th}$ longitudinal mode at a chosen frequency $\overline{\omega}_0$ by solving the following problem for a fixed topology $\overline{\underaccent{\tilde}\chi}$.   

\begin{figure}
\centering
\includegraphics[width=1\textwidth]{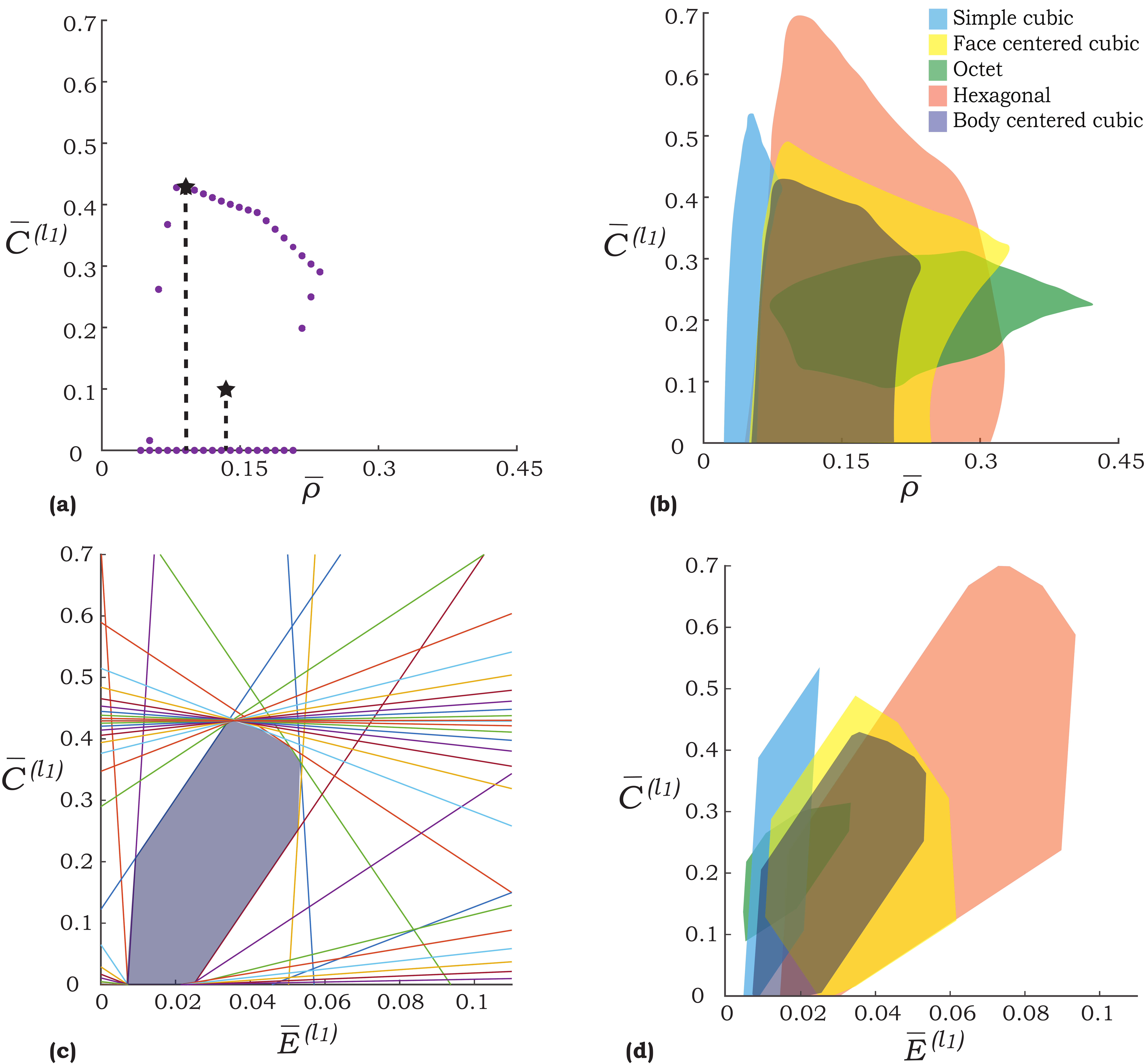}
\caption{\textit{ \textbf{(a)} Construction of bounds on non-dimensional group velocuty vs. non-dimensional density for the BCC lattice at $\overline{\omega}^*=0.35$. \textbf{(b)} Bounds on the non-dimensional group velocity of the chosen longitudinal mode with respect to the non-dimensional density of each lattice. \textbf{(c)} Construction of bounds on non-dimensional group velocity vs. non-dimensional static stiffness for the BCC lattice at $\overline{\omega}^*=0.35$. \textbf{(d)} Bounds on the non-dimensional group velocity of the chosen longitudinal mode with respect to the non-dimensional stiffness of each lattice.}} 
\label{fig:bounds}
\end{figure}

An example showing the construction of bounds is shown in Fig.~\ref{fig:bounds}(a) for the BCC lattice, where we calculate the maximum and minimum group velocities at fixed densities for the range of possible densities. Notice that a minimum wave speed of zero is achieved for a broad range of densities for the chosen longitudinal mode. This indicates that we either find topologies that force $\overline{\omega}^*$ to a band edge or that we observe a standing wave. We pick two topologies of the BCC lattice to experimentally validate our results, one with the fastest propagating wave speed $\left( \overline{C}^{(l_1)}=0.43 \right)$ and the other with an arbitrarily chosen low speed $\left( \overline{C}^{(l_1)}=0.1 \right)$ within the bounds of wave speed, and roughly the same density. The properties of these structures are indicated by dashed lines and star markers in Fig.~\ref{fig:bounds}(a) and the topologies are calculated by solving problem \eqref{eq:opt}. We are limited by the manufacturability of point masses at the nodes, which causes the slight variation in density between the two chosen geometries. In the lattice with the intermediate wave speed, the bars along the diagonal of a unit cell are the thickest ($\overline{A}=17 \times 10^{-3}$) and the bars along the wave propagation vector are the thinnest ($\overline{A}=4.25 \times 10^{-3}$), with point masses at the corners with non-dimensional mass $\overline{m}=5.8 \times 10^{-3}$. In the lattice with the faster wave speed, the bars along the diagonal have the least cross-sectional areas ($\overline{A}=4.25 \times 10^{-3}$) and the bars along the lattice vector have the highest areas ($\overline{A}=17 \times 10^{-3}$) without any point masses at the joints. 
%\begin{equation} \label{mainoptprob_mod_nd}
%\begin{aligned}
%& \underset{\overline{\underaccent{\tilde}\chi}}{\text{min}}
%& & \left( \overline{C}^{(L_n)}\left(\overline{\underaccent{\tilde}\chi},\ \overline{k} \left( %\overline{\underaccent{\tilde}\chi} \right) \right) - \overline{C'} \right)^2\\
%& \text{subject to}
%& & \underaccent{\tilde}B_l \leq \overline{\underaccent{\tilde}\chi} \leq \underaccent{\tilde}B_u \\
%&&& \underaccent{\tilde}A_{eq}. \overline{\underaccent{\tilde}\chi}= \overline{\rho},
%\end{aligned}
%\end{equation}
%where $\overline{C'}$ is the desired non-dimensional wave speed.

Fig.~\ref{fig:bounds}(b) shows the results of the optimization for the five one dimensional periodic lattices of choice at $\overline{\omega}^*=0.35$, in terms of achievable regions of the wave speed vs.\ density space. The hexagonal lattice features the highest wave speeds at low relative densities. This can be attributed to its high static stiffness in the longitudinal direction provided by the six bars aligned along this direction (see Fig.~\ref{fig_2}(e)). Intuitively, placing the thickest bars along the propagation direction in all geometries boosts the long wavelength wave speed at a given density. The SC lattice is the least dense, while still featuring high wave speeds. FCC and BCC lattices allow to span wider density ranges, while presenting similar wave speeds. Finally, the octet lattice presents the lowest maximum wave speeds, owing to the absence of bars aligned in the direction of wave propagation. 

Next, we calculate bounds on the non-dimensional group velocity as a function of the static stiffness, for the five lattices, at $\overline{\omega}^*=0.35$. We maintain the same constraints on the topology of each unit cell as indicated in Table \ref{params}. Once again, to avoid bandgaps  within the bounds of design variables for $\overline{\omega}^*=0.35$, we optimize the group velocity $\overline{C}^{(l_n)}$ of the first longitudinal mode for the SC, BCC, hexagonal and octet lattice, and the third longitudinal mode for the FCC lattice. To calculate the bounds, we solve problem \eqref{eq:optce} using a gradient based method with the sensitivities derived in section \ref{sens_secn}. An example showing the construction of bounds is shown in Fig. \ref{fig:bounds}(c) for the BCC lattice. Here $\overline{E}^{(l_n)} \left( \overline{\underaccent{\tilde}\chi} \right)=E^{(l_n)} \left( \underaccent{\tilde}\chi \right)/E_s$. The shaded region encloses all the attainable values of group velocity and stiffness. We repeat the process for all five lattices to calculate the bounds, as shown in Fig. \ref{fig:bounds}(d). We can achieve a large range of possible wave speeds at fixed stiffness values, even though the two properties are highly dependent and proportional. We are able to decouple these properties because of the presence of point masses, which normally allow for backscattering \cite{wang2015scatt} or resonance \cite{wang2014reson} effects that affect dispersion properties without affecting the static stiffness. However, in our lattices the mode shapes in Fig. \ref{fig_4} do not indicate local resonance at the masses.

%In applications where it is necessary to engineer a combination of wave speed $\overline{C}'$ and static stiffness $\overline{E}'$ at a fixed density $\overline{\rho}$ within the bounds shown in Fig. \ref{fig_3}, we solve the non-dimensional version of problem \eqref{mainoptprob_stiff_mod} described in the previous section.
%\begin{equation} \label{mainoptprob_stiff_mod_nd}
%\begin{aligned}
%& \underset{\overline{\underaccent{\tilde}\chi}}{\text{max}}
%& & \delta_1 \left( \overline{C}^{(L_n)}\left(\overline{\underaccent{\tilde}\chi},\ \overline{k} \left( \overline{\underaccent{\tilde}\chi} \right) \right)- \overline{C}'\right)^2 + \delta_2 \left(\overline{C}_0 \left(\overline{\underaccent{\tilde}\chi} \right)^2 f \left( \overline{\underaccent{\tilde}\chi} \right) - \overline{E}' \right)^2\\
%& \text{subject to}
%& & \underaccent{\tilde}B_l \leq \overline{\underaccent{\tilde}\chi} \leq \underaccent{\tilde}B_u \\
%&&& \underaccent{\tilde}A_{eq}. \overline{\underaccent{\tilde}\chi}= \overline{\rho}.
%\end{aligned}
%\end{equation}

\subsection{Verification with 3-D finite element solver}

\begin{figure}
\centering
\includegraphics[width=1.\textwidth]{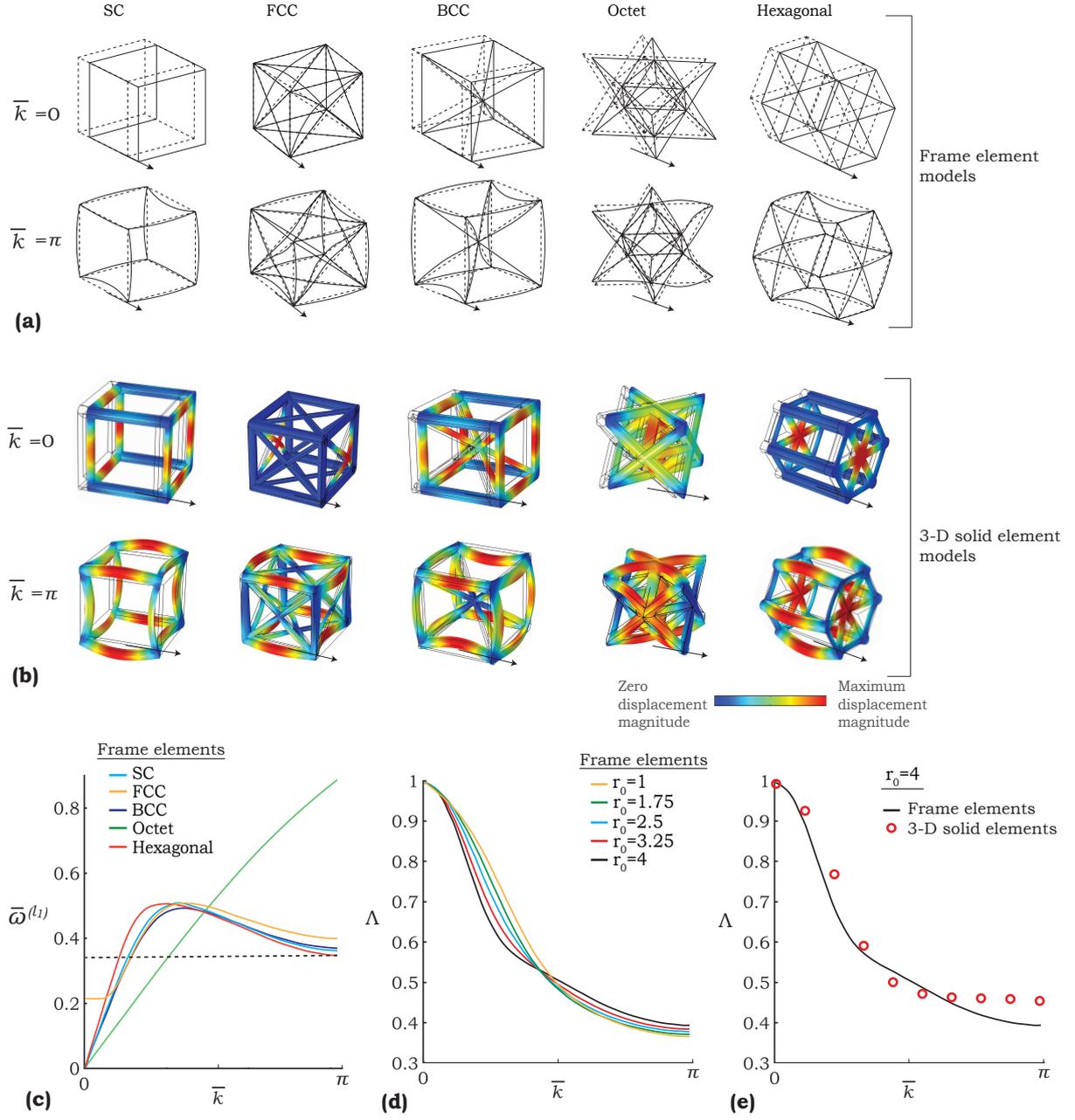}
\caption{\textit{\textbf{(a)} and \textbf{(b)} Mode shapes of the five 1-D periodic lattices obtained by modeling them using frame elements and 3-D solid elements, respectively. \textbf{(c)} Dispersion curves for the topologies with highest wave speed of the longitudinal mode considered, in each of the five 1-D lattices at $\overline{\omega}_0=0.35$. \textbf{(d)} Variation of parameter $\Lambda$ (eq. \eqref{lambda}) of the first longitudinal mode for the 1-D BCC lattice, with the non-dimensional wavenumber for five different ratios of cross-sectional areas. \textbf{(e)} Variation of parameter $\Lambda$ of the first longitudinal mode for the BCC lattice with $r_0=4$ modeled using frame elements and 3-D solid elements. }} 
\label{fig_4}
\end{figure}

We obtain the optimal designs and bounds in section \ref{sec:numbounds} by modeling the bars in each structure using frame elements, which assumes slender bars represented by one dimensional elements. In this subsection, we compare the group velocities and mode shapes of key optimal designs modeled using 3-D solid elements (tetrahedral elements) in COMSOL. Specifically, we compare the wave speeds and mode shapes of topologies that attain peak group velocity in Fig. \ref{fig:bounds}(b) and (d) as these designs are achieved at extremal values of $\overline{\underaccent{\tilde}\chi}$. %, and the wave speeds of the BCC lattice with an intermediate property as shown in Fig. \ref{fig:bounds}(a).
To visualize the cell deformation that is characteristic of the longitudinal modes, we plot mode shapes for each of the five lattices in Fig. \ref{fig_4}(a) and (b) obtained by modeling the structures using frame elements and 3-D solid elements, respectively at extreme wavenumbers. %Note that the structures correspond to the ones that show maximum wave speed within the bounds of design indicated in Fig. \ref{fig_3}. 
Fig. \ref{fig_4}(c) depicts the dispersion curves for the longitudinal modes considered in these lattices modeled using frame elements. Note that the dashed line indicates the frequency of interest, $\overline{\omega}^*=0.35$. 

In order to understand the mode shape of each mode depicted in Fig. \ref{fig_4}(c), we evaluate the mode shape, i.e. the mass normalized eigenvector from eq. \eqref{globaleq} at a given wavenumber $\overline{k}$ for each topology. This eigenvector contains the displacement vector of each node in the lattice  from their initial position in the undeformed state. Note that a mode shape is qualitative in the sense that any scalar multiple of an eigenvector still represents the same mode. We calculate the mode shapes of the five one dimensional lattices at non-dimensional wavenumbers $\overline{k}=0$ and $\pi$, modeled using frame elements (Fig. \ref{fig_4}(a)). We obtain the deformed geometries (indicated by solid curves) after displacing the underformed geometries (indicated by dashed lines) by some scalar multiple of the eigenvector (chosen based on adequate magnification) at each wavenumber. Notice that the mode shape in each geometry involves a rigid body translation of the lattice at the low wavenumber and transitions into a more flexural mode of propagation that involves bending of bars at the high wavenumber. Note that the direction of periodicity of these lattices, which is also the direction of propagation of the wave is indicated by an arrow against each unit cell. In order to verify the mode shapes of propagation, we model the same five lattices in COMSOL with 3-D solid elements. The resulting mode shapes obtained from Bloch wave analyses in these lattices is shown in Fig. \ref{fig_4}(b). Qualitatively comparing figures \ref{fig_4}(a) and (b), is clear that the mode shapes obtained from modeling the lattices as an assembly of bars represented by one dimensional frame elements is almost identical to modeling the whole lattice using 3-D solid elements.

 In order to verify the mode shapes quantitatively, we take a look at a non-dimensional parameter that is a characteristic of the mode shape, and its correlation with longitudinal wave speed and mode shape. We have that, the mass normalized eigenvector obtained from eq.~\eqref{globaleq} for a given topology and wavenumber is $\overline{\underaccent{\tilde}U}$. We can define constant projection matrices $\underaccent{\tilde}P_1$, $\underaccent{\tilde}P_2$ and $\underaccent{\tilde}P_3$, so that the non-zero components of the vectors $\underaccent{\tilde}P_1 \overline{\underaccent{\tilde}U}$, $\underaccent{\tilde}P_2 \overline{\underaccent{\tilde}U}$ and $\underaccent{\tilde}P_3 \overline{\underaccent{\tilde}U}$ indicate the non-zero non-dimensional displacements of the nodes along three mutually perpendicular directions with $\underaccent{\tilde}P_1 \overline{\underaccent{\tilde}U}$ along the direction of wave propagation. Now, we define a ratio

\begin{equation} \label{lambda}
   \Lambda= \frac{\sqrt{\underaccent{\tilde}P_1 \overline{\underaccent{\tilde}U}.\underaccent{\tilde}P_1 \overline{\underaccent{\tilde}U}}}{\sum_i \sqrt{\underaccent{\tilde}P_i \overline{\underaccent{\tilde}U}.\underaccent{\tilde}P_i \overline{\underaccent{\tilde}U}}}, 
\end{equation}
which is a non-dimensional measure of the longitudinal displacement (i.e., the displacement along the direction of propagation of the wave) relative to the total displacement of all nodes for the mode under consideration at a given topology and wavenumber of propagation. To study the evolution of this ratio with topology, we pick as an example the one dimensional BCC lattice with different topologies by varying the area of the diagonal bars. Let $r_0$ denote the ratio of the non-diagonal bars' cross-sectional area to the diagonal bars' cross-sectional area, with the thickest possible area along the non-diagonal bars. For every topology, we notice that the ratio $\Lambda$ starts at one (i.e., all displacements are along the direction of propagation of the wave) at $\overline{k}=0$ and decays to values that are less than half at $\overline{k}=\pi$, as shown in Fig. \ref{fig_4}(d). This indicates that majority of displacements at $\overline{k}=\pi$ are perpendicular to the direction of wave propagation. The behavior is consistent with the qualitative mode shapes indicated in figures \ref{fig_4}(a) and (b). To quantitatively compare the mode shapes of the lattices modeled using frame elements and 3-D solid elements, we compute the parameter $\Lambda$ for the BCC lattice with $r_0=4$ using both models. The results are shown in Fig. \ref{fig_4}(e). The mean squared error, calculated between the data points indicated by red circles (3-D solid elements) and the black curve (frame elements) is $0.67\%$, further indicating that both models display similar longitudinal mode shapes.

In Table \ref{comparison_solvers} we compare the non-dimensional group velocities associated with the topologies of the five lattices with maximum wave speed (Fig. \ref{fig:bounds}(b) and (d)) modeled using both methods. We see great agreement in measured wave speeds with the two methods as the bounds on the topological vector ensure that the bars remain slender for the most part. The discrepancies mainly occur when the thickest bars are placed between joints that are much closer to each other, reducing their slenderness (ratio of length to diameter), which is often the case to obtain maximum wave speeds in these lattices. For the topology of the BCC lattice with an intermediate wave speed (Fig. \ref{fig:bounds}(a)), the non-dimensional wave speed calculated with a model using frame elements is 0.10, while the wave speed of the lattice modeled using 3-D solid elements is 0.24. This discrepancy can be attributed to a low slenderness ratio of 5.88 of the diagonal bars in the optimal design, which implies the experiments in the next section must be verified with the 3-D solid element model over the frame element model for accuracy.  %The discrepancies in wave speeds are further examined in Fig. \ref{fig_5}.

\begin{table}  
  \begin{center}
    \caption{Comparison of non-dimensional group velocities of lattices that attain maximum $\overline{C}^{(l_n)}$ in Fig. \ref{fig:bounds}(b)}
   \label{comparison_solvers}
    \begin{tabular}{c|c|c} 
      \textbf{Lattice} &\textbf{1-D frame elements}& \textbf{3-D solid elements} \\
      \hline
      \rule{0pt}{3ex}
      Hexagonal & 0.70 & 0.81 \\ \rule{0pt}{3ex}
      SC  & 0.54 & 0.57\\ \rule{0pt}{3ex}
      BCC  & 0.43 & 0.50 \\ \rule{0pt}{3ex}
      FCC & 0.48 & 0.46 \\  \rule{0pt}{3ex} 
      Octet & 0.32 & 0.40 \rule{0pt}{3ex} 
    \end{tabular}
  \end{center}
\end{table}

\begin{figure}
\centering
\includegraphics[width=0.9\textwidth]{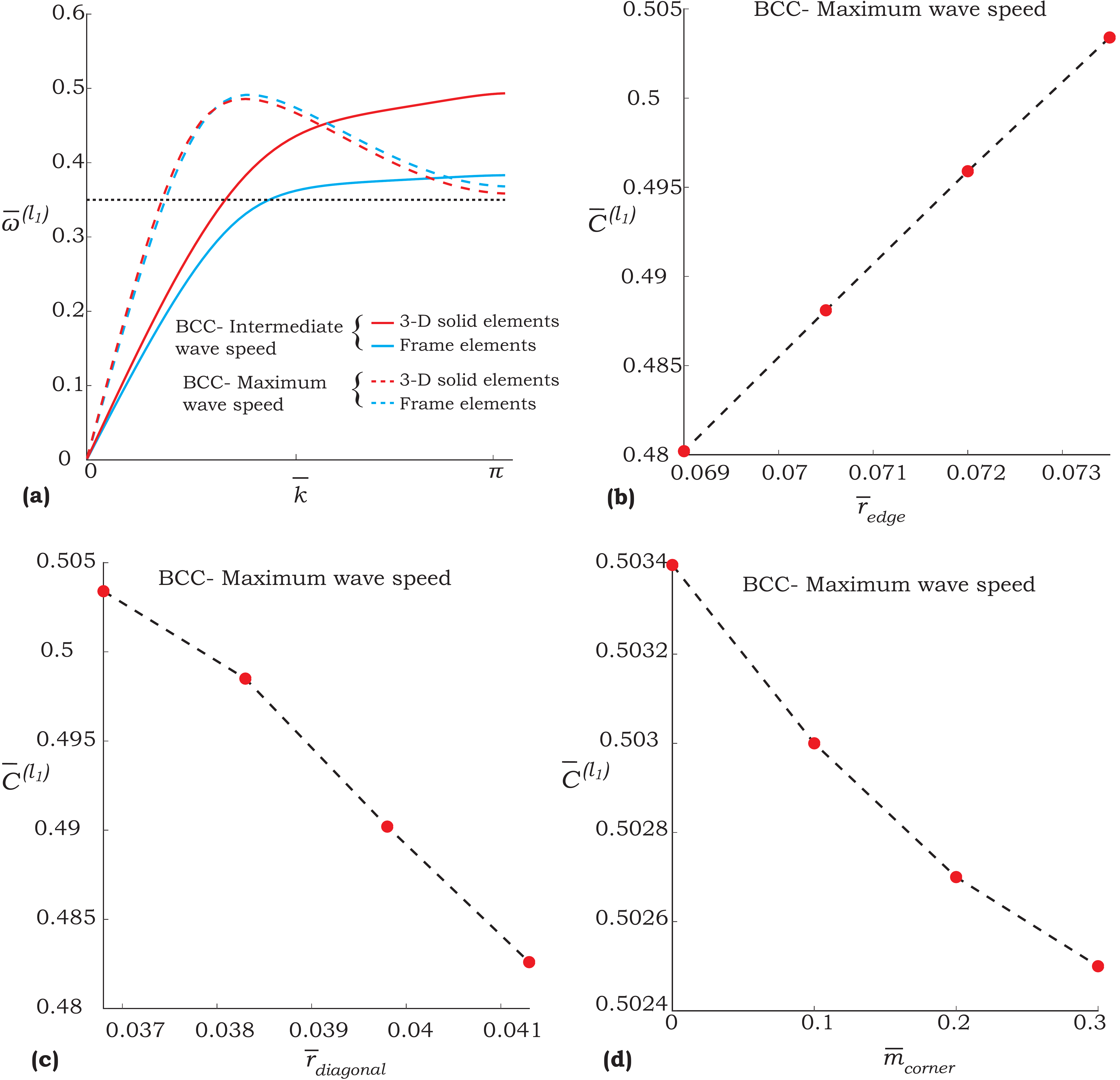}
\caption{\textit{\textbf{(a)} Dispersion curves obtained by modeling the BCC lattice with maximum and intermediate wave speeds, using frame elements and 3-D solid elements. \textbf{(b)}, \textbf{(c)} and \textbf{(d)} Variation of wave speed of BCC lattice with maximum wave speed, with the radius of its edge bars, radius of its diagonal bars and masses at the corners of a unit cell, respectively modeled using 3-D solid elements. }} 
\label{fig_5}
\end{figure}

 In Fig. \ref{fig_5}(a) we show the dispersion curve of the first longitudinal mode plotted using both models for the topologies of BCC lattice with maximum and intermediate wave speed mentioned in the previous section. While the dispersion curves for the lattice with maximum wave speed are nearly identical (dashed curves), the difference in curves for the lattice with an intermediate speed (solid curves) mostly occurs at higher wave numbers and frequencies, where the lattice is more dispersive. As a result, we see differences in wave speeds between the two methods at larger frequencies such as $\omega^*=0.35$ (black dotted line). %For the lattice with intermediate wave speed, the thickest bars are along the body diagonal and the thinnest bars are along the edges, with point masses at the corners. The 3-D model depicting this topology is shown in the inset in Fig. \ref{fig_5}(b). Notice the masses are modeled as spheres with higher density ($\overline{\rho}=16.02$) to match the value of the point mass described in the previous section and best replicate the experiments described in the following section.
 We examine the sensitivity of topology variables on the wave speed in the lattice with maximum wave speed. For this lattice, the thickest bars are along the edges while the thinnest bars are along the body diagonal.  %For the topology that corresponds to an intermediate wave speed, it can be shown that the sensitivity of the wave speed with respect to the cross-sectional area of the edge bars and diagonal bars is positive according to eq. \eqref{sensa} when modeled using frame elements. These trends can be verified for the topology modeled using 3-D solid elements, as shown in Fig. \ref{fig_5} (b) and (c). Notice that the wave speed increases with increasing radius of the edge bars ($\overline{r}_{edge}$) and decreases with decreasing radius of the diagonal bars ($\overline{r}_{diagonal}$). 
When modeled using frame elements, the sensitivities of the wave speed with respect to the cross-sectional areas of edge bars, diagonal bars and masses at corners ($\overline{m}_{corner}$) are positive, negative and negative, respectively according to eq. \eqref{sensa}. We verify these trends and ensure that the topology corresponds to a local maximum in wave speed when modeled using 3-D solid elements, as shown in Fig. \ref{fig_5}(b), (c) and (d). This shows that the extremal topology still remain optimal or displays similar trends in wave speed with respect to topological parameters regardless of the method chosen to model.

\section{Experiments} \label{exp}
In this section, we explain how our samples are designed and fabricated, and we illustrate a procedure to excite and study the spectro-spatial characteristics of the longitudinal modes. We then measure the wave speed of the first longitudinal mode at a particular frequency in the samples and compare it to the wave speeds computed theoretically. We also experimentally reconstruct the dispersion curves for the selected mode around the frequency of operation, and compare them to our finite element calculations.

\subsection{Design of structures with engineered group velocities}

\begin{figure}
\centering
\includegraphics[width=.75\textwidth]{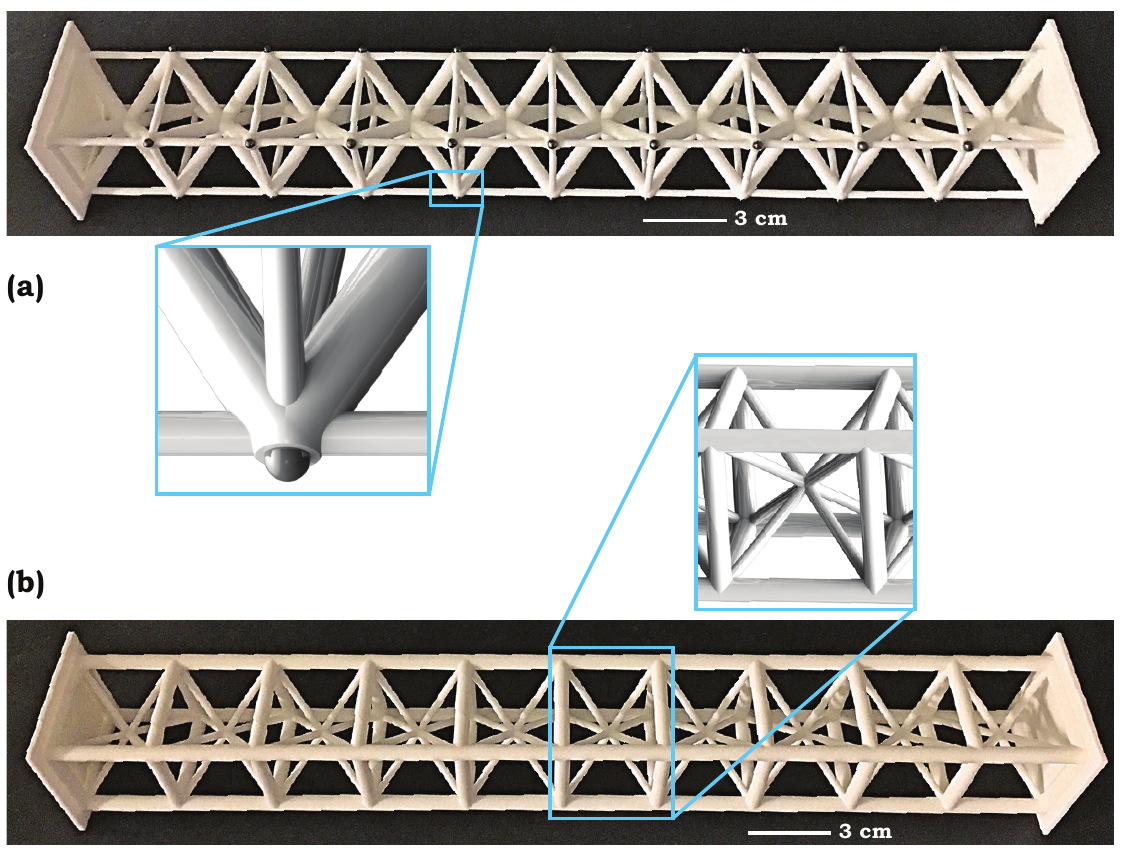}
\caption{\textit{\textbf{(a)} BCC lattice with an intermediate wave speed (featuring 3 mm diameter point masses at the joints, as illustrated in the inset). \textbf{(b)} BCC lattice with the highest wave speed (the inset illustrates a closeup view of the unit cell).}} 
\label{fig_6}
\end{figure}

We focus our attention on two samples with properties indicated by the dashed lines in Fig. \ref{fig:bounds}(a), as these lattices exhibit similar densities but vastly different wave speeds. The optimal values for both topologies are obtained from problem \eqref{eq:opt} and detailed in section \ref{sec:numbounds}. We then pick the closest experimentally realisable topologies (Fig.~\ref{fig_6}). The first corresponds to the BCC lattice with the highest wave speed of 0.50 at $\overline{\omega}^*=0.35$, and the second has an engineered non-dimensional wave speed of 0.24 at the same frequency, when modeled using 3-D solid elements.  

In order to experimentally test the properties of these lattices, we fix our frequency of operation at 2200 Hz ($\omega^*=13.82$ krad/s). This frequency gives us a lattice parameter (unit cell length, $L$) of 34 mm at $\overline{\omega}^*=0.35$ according to the relation $\overline{\omega}^*= \omega^* \sqrt{\frac{\rho_s}{E_s}} L$. We choose this frequency as it yields lattices within a convenient length scale to fabricate and test. However, as mentioned in the previous sections, the designs can be scaled to any frequency of operation as the analysis is non-dimensionalised. A 3-D printed lattice at this length scale can be conveniently manufactured using a selective laser sintering process with our chosen material (Polyamide 12). The properties of this material are summarized in Table \ref{params}. As previously mentioned, in the lattice with the intermediate wave speed, the bars along the diagonal of a unit cell have non-dimensional area $\overline{A}=17 \times 10^{-3}$ (with area 19.65 mm$^2$) and the bars along the direction of periodicity have non-dimensional area $\overline{A}=4.25 \times 10^{-3}$ (with area 4.91 mm$^2$), with point masses at the corners with non-dimensional mass $\overline{m}=5.8 \times 10^{-3}$ (tungsten carbide ball bearings with diameter 3 mm as seen in Fig.~\ref{fig_6}(a)). In the lattice with the faster wave speed, the bars along the diagonal have non-dimensional area $\overline{A}=4.25 \times 10^{-3}$ (with area 4.91 mm$^2$)  and the bars along the lattice vector have the non-dimensional area $\overline{A}=17 \times 10^{-3}$ (with area 19.65 mm$^2$) without any point masses at the joints  (Fig.~\ref{fig_6}(b)).

%In order to experimentally test the properties of these lattices, we fix our frequency of operation at 2200 Hz ($\omega^*=13817$ rad/s). This frequency gives us a lattice parameter (unit cell length, $L$) of 34 mm at $\overline{\omega}^*=0.35$ according to the relation $\overline{\omega}^*= \omega^* \sqrt{\frac{\rho_s}{E_s}} L$. We choose this frequency as it yields lattices within a convenient length scale to fabricate and test. However, as mentioned in the previous sections, the designs can be scaled to any frequency of operation as the analysis is non-dimensionalised. A 3-D printed lattice at this length scale can be conveniently manufactured using a selective laser sintering process with our chosen material (Polyamide 12). The properties of this material are summarized in Table \ref{params}. 

The lattice with intermediate wave speed (Fig.~\ref{fig_6}(a)) features non-zero point masses at the cell corners. We realize these experimentally by press-fitting 3 mm diameter ball bearings made of tungsten carbide into cavities that are designed at the corners of each unit cell, as shown in Fig.~\ref{fig_6}(a). The ball bearings are over 16 times more dense (density of $14900\,\mathrm{kg/m^3}$) than the solid material that the lattice frame is made of. The fact that we have a relatively high mass within a small volume at the joints supports our assumption of point masses. The sample with the the highest wave speed (Fig. \ref{fig_6}(b)) does not include point masses at the joints and does not need any special assembly after fabrication. As mentioned earlier, we slightly relax the constant density constraint, as the sample fabrication is limited by the ball bearings' availability. We fabricate two sets of samples: arrays of 10 units and arrays of 15 units. The 10-unit arrays are used to measure wave speeds, while the longer ones are used for dispersion reconstruction. 

\subsection{Setup}

\begin{figure}
\centering
\includegraphics[width=1.\textwidth]{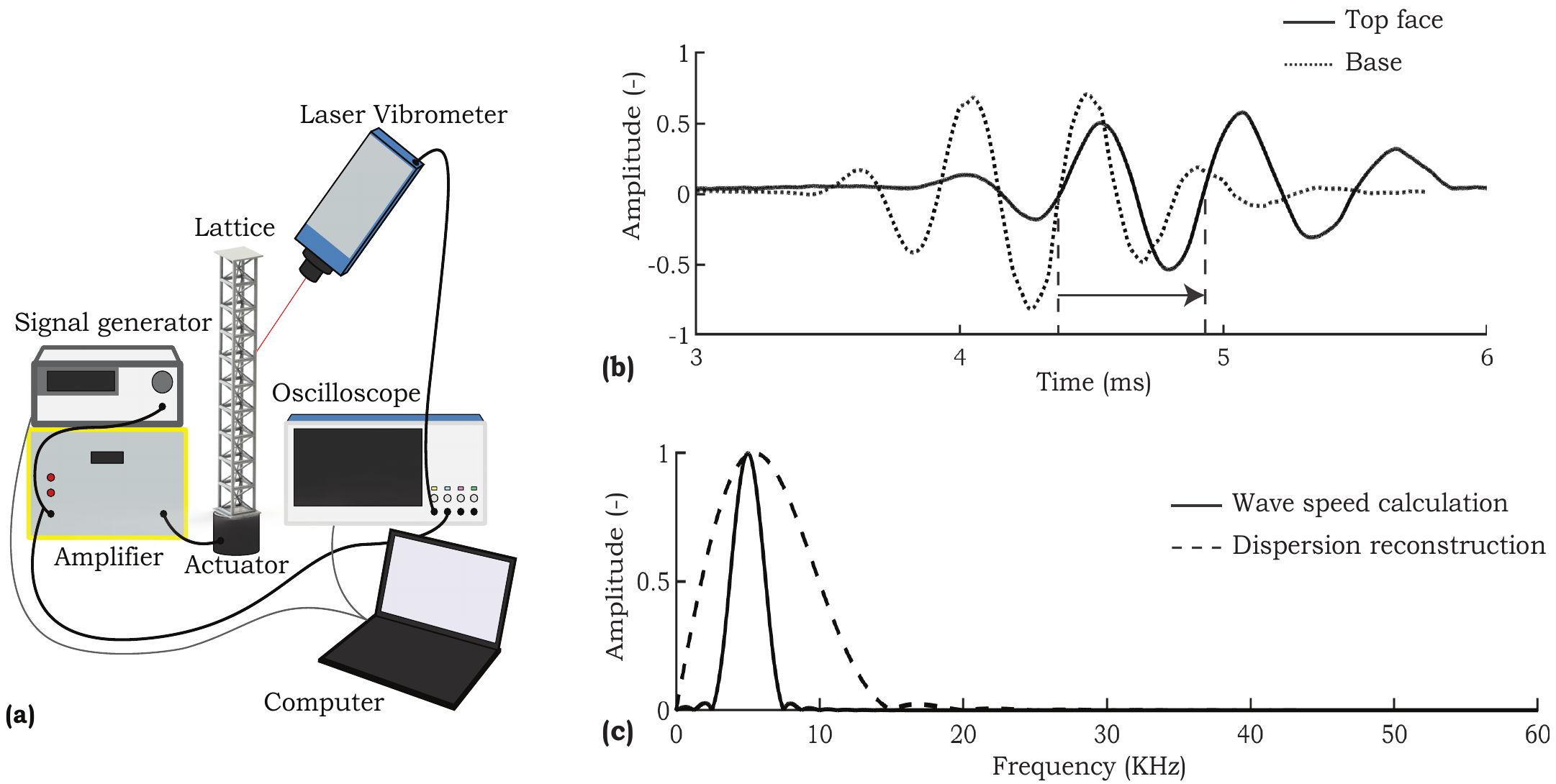}
\caption{\textit{\textbf{(a)} Layout of the experimental setup used to measure the elastic wave response of the lattices.
\textbf{(b)} Example of a measured transient wave response (normalized by the maximum positive amplitude of the input signal) at two points of the lattice. \textbf{(c)} Frequency response of the input signals used to calculate wave speed and reconstruct the dispersion structure.}} 
\label{fig_7}
\end{figure}

The experimental setup is sketched in Fig.~\ref{fig_7}(a). To measure the group velocity through a sample at the frequency $\omega^*$, we construct a burst input signal with four cycles and measure the time taken for the centroid of the wave packet to travel through the sample, as shown with the signals in Fig.~\ref{fig_7}(b). Note that the amplitude of each signal is normalized by the maximum positive amplitude of the input signal. This wave packet is narrow-banded around the frequency of interest, as indicated by the solid curve in Fig. \ref{fig_7}(c), which is obtained from the DFT of the input signal. The signal is communicated to the signal generator (Agilent 33220A), sent to a piezoelectric amplifier (Piezomechanik LE 150/100 EBW) and to a piezoelectric transducer (Panametrics V1011). The transducer imparts a longitudinal excitation to the base of the lattice array. This ensures that no other modes (shear or torsional) are excited in the lattice. We then use a laser doppler vibrometer (LDV, Polytec OFV-5000) to measure the signal at the base and at the top of the specimen. The signals are read by an oscilloscope (Tektronix DPO3014) and fed back to a computer. The wave speed is then calculated from the signal's time-of-flight between measurement points. We choose arrays of 10 units for the wave speed measurements since this dimension corresponds to over five wavelengths of the longitudinal wave at the frequency of interest. This guarantees separation between the incoming and reflected wave packets when we measure at the top of the array (Fig. \ref{fig_7}(b)). To reconstruct the dispersion properties of the lattice, we instead use a broadband signal (a burst with carrier $f_0$ but featuring a single oscillation), The wave packet is broader around the frequency of interest, as indicated by the dashed curve in Fig. \ref{fig_7}(c). We then record the response on the bars that are perpendicular to the direction of wave propagation. The time histories at all measurement points are then collected into a matrix. Upon 2D Discrete Fourier Transform (2D-DFT), we obtain a frequency-wavenumber matrix, containing information on the dispersion characteristics of a truncated lattice, from the time-space data~\cite{Palermo2019}. In this case, we choose arrays of 15 cells since our setup only allows one measurement per cell plus one at the top of the lattice, and 16 measurement points guarantee enough spatial resolution.

\subsection{Results}

We record the burst input signal with four cycles at four corners of each sample on the base and its top face. From this, we calculate sixteen wave speeds of the longitudinal mode through a sample. Finally, we compute the group velocity as an average of these sixteen speeds. We repeat this over two samples for each of the BCC lattices, i.e., we obtain the group velocity as the average of 32 wave speeds measured across different points on different samples for each lattice of the types shown in Fig.~\ref{fig_6}. We non-dimensionalise the group velocity by dividing the measured speed by $\sqrt{\frac{E_s}{\rho_s}}$ and compare it to our finite element calculations. These results are summarized in Table \ref{results}.

\begin{table}  
  \begin{center}
    \caption{Comparison of non-dimensional (dimensional) group velocities between experiments and numerical calculations}
   \label{results}
    \begin{tabular}{c|c|c} 
      \textbf{Method} &\textbf{Fig. \ref{fig_6}(a)}& \textbf{Fig. \ref{fig_6}(b)}\\
      \hline
      \rule{0pt}{3ex}
      %Frame element model & 0.100 & 0.491 \\ \rule{0pt}{3ex}
      Experiments  & $0.19 \pm 0.02$ $(256.88 \pm 27.04\ \textrm{m/s})$ & $0.49 \pm 0.03$ $(662.48 \pm 40.56\ \textrm{m/s})$ \\ \rule{0pt}{3ex}
      3-D solid elements model  & 0.24 (324.48 m/s) & 0.50 (681.40 m/s) \\ %\rule{0pt}{3ex}
    \end{tabular}
  \end{center}
\end{table}

We notice that the numerical and experimental results agree well for the sample with maximum wave speed (Fig. \ref{fig_6}(b)). We observe a considerable decrease in wave speed on introducing masses (Fig. \ref{fig_6}(a)) at the joints, as verified by the solid elements model. The slight differences between experiments and numerics can occur due to the fact that the input wave, although concentrated at the frequency of interest, has a slight frequency spread as seen in Fig. \ref{fig_7}(c). Note that the experimental results show that we can obtain more than a 160\% increase in wave speed between the two BCC samples with nearly the same densities---an aspect that can have interesting practical implications.   

\subsection{Dispersion reconstruction}

\begin{figure}
\centering
\includegraphics[width=1.\textwidth]{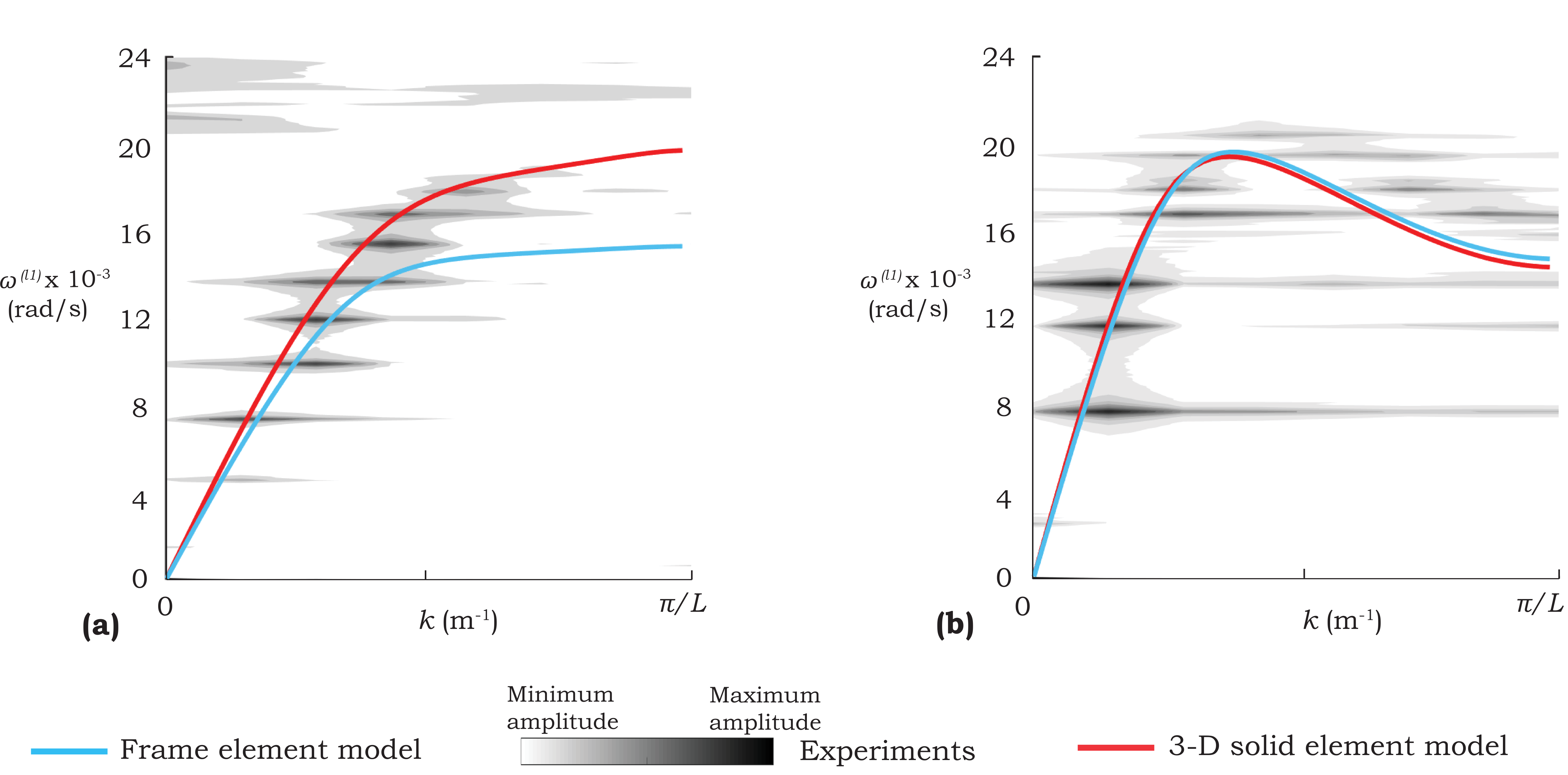}
\caption{\textit{\textbf{(a)} and \textbf{(b)} Experimental dispersion reconstruction and comparison of the first longitudinal branch for the BCC lattices in figures \ref{fig_6}(a) and (b), respectively.}} 
\label{fig_8}
\end{figure}

To understand if we are indeed measuring experimental wave speeds that correspond to the desired longitudinal mode, we experimentally reconstruct the dispersion relation around the frequency of interest $\omega^*$. Choosing an input burst with just one cycle guarantees that the DFT of this signal has a wide frequency spread around $\omega^*$ as indicated by the dashed curve in Fig.~\ref{fig_7}(c). This input signal is used to excite the longer samples and, as explained above, the output signal (amplitude as a function of time) is measured at sixteen equally spaced points starting from the base of the sample to its top. The colormaps in Fig.~\ref{fig_8} represent the reconstructed frequency vs.\ wavenumber plots for the two lattices tested. The darkest regions represent the highest amplitudes at a particular frequency and wavenumber. One can notice that some frequencies have higher amplitude than others. This originates from the fact that, due to the limited length of the samples, the recorded signals include several boundary reflections and, therefore, contain some signature of the natural frequencies of the sample~\cite{Celli2019}.  We compare these color-maps to the first longitudinal branches obtained using both a frame element model (blue curve) and 3-D solid element model with COMSOL (red curve). The numerical results roughly overlap the maxima of the experimental colormaps, indicating agreement between the dispersion for an infinite lattice and the experimental results on a finite sample. The discrepancy in Fig.~\ref{fig_8}(a), where the experiments match the results obtained by modeling the lattice using 3-D solid elements better than the model that uses frame elements, is consistent with arguments discussed in the previous section.

\section{Conclusion}
We present a sensitivity analysis to optimize the elastic group velocity of a structure at a fixed frequency of propagation together with static properties, such as mass density and static stiffness along any chosen direction. In calculating the sensitivities, we provide a method to compute the implicit derivatives, i.e., the sensitivities of the eigenvectors describing the mode shape, with respect to the topology of the structure. We then present a design method to optimize the elastic group velocity in lattice structures, independently from the static properties, density and stiffness. We are able to decouple these properties by adding point masses to the structure at specific locations. We target longitudinal modes in five common lattice structures (SC, BCC, FCC, octet and hexagonal) and provide bounds on the wave speed with the density and static stiffness of each structure while varying cross-sectional areas of the bars and masses at their joints. The bounds are calculated in the structures by modeling them as an assembly of bars made of several frame elements. We then verify the theoretical model by comparing group velocities and mode shapes of representative structures modeled using 3-D solid elements. To experimentally validate our results, we test two 3-D printed BCC lattices, one with the highest group velocity within our chosen bounds and another with an intermediate wave speed. To verify that the experimentally calculated wave speeds correspond to the desired longitudinal mode, we reconstruct the dispersion relation around the frequency of operation and compare it to the numerically simulated curves. 

The methods presented in this paper can be easily extended to optimize the wave speeds of other modes (e.g., shear or torsional) displayed by lattice structures. The non-dimensionalisation of our analysis allows us to perform the optimization at a fixed non-dimensional frequency that scales with a chosen length dimension of the lattice for a given frequency of operation. Finally, the analysis in the paper can be used to design stiff and lightweight structures with arbitrary dispersion characteristics.

\section*{Acknowledgements}
This work was supported by the Shang-Li and Betty Huang endowed graduate fellowship fund in mechanical engineering at the California Institute of Technology. CD and PC acknowledge support from the National Science Foundation (NSF) CSSI grant number 1835735.  We thank Lorenz Affentranger for helping us with the experimental setup.

\section*{Appendix A}
The non-dimensional elemental stiffness matrix $\overline{\underaccent{\tilde}K}_{el}$ and mass matrix $\overline{\underaccent{\tilde}M}_{el}$ are $12\times 12$ symmetric matrices with non-zero elements in their upper triangular portions being
\begin{align*}
  &\overline{\underaccent{\tilde}K}_{el(1,1)}= \overline{A}n,\ \overline{\underaccent{\tilde}K}_{el(1,7)}= -\overline{A}n,\ \overline{\underaccent{\tilde}M}_{el(1,1)}= \frac{2}{3}\overline{A} \overline{a},\ \overline{\underaccent{\tilde}M}_{el(1,7)}= \frac{1}{3}\overline{A} \overline{a}\\
 &\overline{\underaccent{\tilde}K}_{el(2,2)}= 12\overline{I}n^3,\ \overline{\underaccent{\tilde}K}_{el(2,6)}= 6\overline{I}n^2,\ \overline{\underaccent{\tilde}K}_{el(2,8)}=-12\overline{I}n^3,\ \overline{\underaccent{\tilde}K}_{el(2,12)}= 6\overline{I}n^2, \\  &\overline{\underaccent{\tilde}M}_{el(2,2)}= \frac{78}{105}\overline{A} \overline{a},\ \overline{\underaccent{\tilde}M}_{el(2,6)}=  \frac{22}{105}\overline{A} \overline{a}^2,\ \overline{\underaccent{\tilde}M}_{el(2,8)}=\frac{27}{105}\overline{A} \overline{a},\ \overline{\underaccent{\tilde}M}_{el(2,12)}= -\frac{13}{105}\overline{A} \overline{a}^2, \\
 &\overline{\underaccent{\tilde}K}_{el(3,3)}= 12\overline{I}n^3,\ \overline{\underaccent{\tilde}K}_{el(3,5)}=-6\overline{I}n^2,\ \overline{\underaccent{\tilde}K}_{el(3,9)}=-12\overline{I}n^3,\ \overline{\underaccent{\tilde}K}_{el(3,11)}=-6\overline{I}n^2, \\  &\overline{\underaccent{\tilde}M}_{el(3,3)}= \frac{78}{105}\overline{A} \overline{a},\ \overline{\underaccent{\tilde}M}_{el(3,5)}=  -\frac{22}{105}\overline{A} \overline{a}^2,\ \overline{\underaccent{\tilde}M}_{el(3,9)}=\frac{27}{105}\overline{A} \overline{a},\ \overline{\underaccent{\tilde}M}_{el(3,11)}= \frac{13}{105}\overline{A} \overline{a}^2, \\
 &\overline{\underaccent{\tilde}K}_{el(4,4)}= \frac{G_s}{E_s}\overline{J}n,\ \overline{\underaccent{\tilde}K}_{el(4,10)}= -\frac{G_s}{E_s}\overline{J}n,\ \overline{\underaccent{\tilde}M}_{el(4,4)}= \frac{2}{3}\overline{A} \overline{a} \overline{r}^2,\ \overline{\underaccent{\tilde}M}_{el(4,10)}= -\frac{1}{3}\overline{A} \overline{a} \overline{r}^2 \\
 &\overline{\underaccent{\tilde}K}_{el(5,5)}= 4\overline{I}n,\ \overline{\underaccent{\tilde}K}_{el(5,9)}=6\overline{I}n^2,\ \overline{\underaccent{\tilde}K}_{el(5,11)}=2\overline{I}n,\ \\
 &\overline{\underaccent{\tilde}M}_{el(5,5)}= \frac{8}{105}\overline{A} \overline{a}^3,\ \overline{\underaccent{\tilde}M}_{el(5,9)}=-\frac{13}{105}\overline{A} \overline{a}^2,\ \overline{\underaccent{\tilde}M}_{el(5,11)}=-\frac{6}{105}\overline{A} \overline{a}^3, \\
  &\overline{\underaccent{\tilde}K}_{el(6,6)}= 4\overline{I}n,\ \overline{\underaccent{\tilde}K}_{el(6,8)}=-6\overline{I}n^2,\ \overline{\underaccent{\tilde}K}_{el(6,12)}=2\overline{I}n,\ \\
 &\overline{\underaccent{\tilde}M}_{el(6,6)}= \frac{8}{105}\overline{A} \overline{a}^3,\ \overline{\underaccent{\tilde}M}_{el(6,8)}=\frac{13}{105}\overline{A} \overline{a}^2,\ \overline{\underaccent{\tilde}M}_{el(6,12)}=-\frac{6}{105}\overline{A} \overline{a}^3, \\
 &\overline{\underaccent{\tilde}K}_{el(7,7)}= \overline{A}n,\ \overline{\underaccent{\tilde}M}_{el(7,7)}= \frac{2}{3}\overline{A} \overline{a},\\
 &\overline{\underaccent{\tilde}K}_{el(8,8)}=  12\overline{I}n^3,\ \overline{\underaccent{\tilde}K}_{el(8,12)}= - 6\overline{I}n^2,\ \overline{\underaccent{\tilde}M}_{el(8,8)}= \frac{78}{105}\overline{A} \overline{a},\
 \overline{\underaccent{\tilde}M}_{el(8,12)}= -\frac{22}{105}\overline{A} \overline{a}^2\\
 &\overline{\underaccent{\tilde}K}_{el(9,9)}=  12\overline{I}n^3,\ \overline{\underaccent{\tilde}K}_{el(9,11)}=  6\overline{I}n^2,\ \overline{\underaccent{\tilde}M}_{el(9,9)}= \frac{78}{105}\overline{A} \overline{a},\ \overline{\underaccent{\tilde}M}_{el(9,11)}= \frac{22}{105}\overline{A} \overline{a}^2\\
 &\overline{\underaccent{\tilde}K}_{el(10,10)}= \frac{G_s}{E_s}\overline{J}n,\ \overline{\underaccent{\tilde}M}_{el(10,10)}= \frac{2}{3}\overline{A} \overline{a} \overline{r}^2,\\
  &\overline{\underaccent{\tilde}K}_{el(11,11)}= 4\overline{I}n,\ \overline{\underaccent{\tilde}M}_{el(11,11)}= \frac{8}{105}\overline{A} \overline{a}^3,\\
  &\overline{\underaccent{\tilde}K}_{el(12,12)}= 4\overline{I}n,\ \overline{\underaccent{\tilde}M}_{el(12,12)}= \frac{8}{105}\overline{A} \overline{a}^3,
\end{align*}

where $n$ is the number of elements in the bar, $L$ is the length of the lattice vector, $\overline{A}=\frac{A}{L^2}$ with $A$ being the circular cross-sectional area of the beam and $\overline{a}=\frac{L_{bar}}{2nL}$. $L_{bar}$ is the total length of the bar that the element is a part of. $\overline{I}=\frac{\overline{A}^2}{4\pi}$, $\overline{r}=\frac{\overline{A}}{2\pi}$ and $\overline{J}=2 \overline{I}$ are the non-dimensional planar second moment of area, radius of gyration and polar second moment of area, respectively. $G_s$, the shear modulus of rigidity can be calculated as $\frac{E_s}{2(1+\nu)}$ for the linear elastic material, where $E_s$ and $\nu$ are the Young's modulus of elasticity and Poisson's ratio, respectively. %Finally, the equilibrium equations for an element take the form
%\begin{equation*}
%    \overline{\underaccent{\tilde}M}_{el} \ddot{\overline{\underaccent{\tilde}u}}_{el} + \overline{\underaccent{\tilde}K}_{el} \overline{\underaccent{\tilde}u}_{el} =\overline{\underaccent{\tilde}F}_{el}.
%\end{equation*}

\section*{Appendix B}

\begin{algorithm}[H]
\caption{Non-dimensional frequency of propagation $\overline{\omega}^{(l_n)}$ of the $n^{th}$ longitudinal mode for a non-dimensional wavenumber $\overline{k}$}\label{alg:alg1}
\begin{algorithmic}[1]
\State For the fixed topology $\overline{\underaccent{\tilde}\chi}$ of the unit cell and $\overline{k}$, find the eigenvalues and associated eigenvectors of $\overline{\underaccent{\tilde}M}^{-1} \overline{\underaccent{\tilde}K}$ ($\overline{\omega}_i^2$ and $\overline{\underaccent{\tilde}U}_i,\ i=1,\dots,size\left(\overline{\underaccent{\tilde}M}\right)$)
\State Modify $\overline{\underaccent{\tilde}U}_i$ by subtracting from the displacement of each node, the displacement of the centroid of the unit cell
\State Compute the direction cosines of each bar before displacement, $\underaccent{\tilde}d_j,\ j=1,\dots,$ number of bars
\State Denote the set of indices of bars on the outer frame of the front face of the unit cell by S1
\State Denote the set of indices of bars along $\hat{\underaccent{\tilde}k}$ (direction of wave vector) by S2 and perpendicular $\hat{\underaccent{\tilde}k}$ by S3. $\hat{(.)}$ represents a unit vector.
\State count=0
\For{$i=1:1:size\left(\overline{\underaccent{\tilde}M}\right)$}                    
        \State {compute the direction cosines of the line joining the displaced end points of each bar,                        $\underaccent{\tilde}e_j$ and the mean displacement of nodes along each bar $\underaccent{\tilde}v_j$, $j=1,\dots,$ number of bars}
        \State $p=1$
        \For{$k=1:1:size(S1)$}
            \State {$p=p\times |\underaccent{\tilde}d_{S1(k)}.\underaccent{\tilde}e_{S1(k)}| $} (Indicates a shear/ torsional distortion of lattice if $p<1$)
            \EndFor \State \textbf{end for}
        \State $q=1$
        \For{$k=1:1:size(S2)$}
            \State {$q=q\times \left(1-|\hat{\underaccent{\tilde}k}.\hat{\underaccent{\tilde}v}_j|\right) $}
            \EndFor \State \textbf{end for}    
        \State $t=1$
        \For{$k=1:1:size(S3)$}
            \State {$t=t\times |\hat{\underaccent{\tilde}k}.\hat{\underaccent{\tilde}v}_j| $}
            \EndFor \State \textbf{end for}  
         \State $s=p \times q \times t$   (s=1 for a longitudinal mode)
         
         \State \textbf{if} $s==1$
         \State \quad count=count+1
         \State \quad \textbf{if} count$==n$
         \State \quad  \quad break all
         \State \quad \textbf{end if}
         \State \textbf{end if}

    \EndFor \State \textbf{end for}
    \State $\overline{\omega}^{(l_n)}=\overline{\omega}_i$
\end{algorithmic}
\end{algorithm}

\bibliographystyle{unsrtnat}
\bibliography{bib_main}

\end{document}